\documentclass[superscriptaddress,aps,preprintnumbers,amsmath,amssymb,prd,nofootinbib,preprint]{revtex4-1}
\pdfoutput=1

\usepackage{here}
\usepackage{comment,braket}
\usepackage{epsf}
\usepackage{amsmath}
\usepackage{graphics}
\usepackage{amsfonts}
\usepackage{amssymb}
\usepackage{latexsym}
\usepackage{color}
\input{colordvi.tex}
\usepackage{feynmp}

\numberwithin{equation}{section}
\renewcommand{\thesection}{\arabic{section}}
\renewcommand{\thesubsection}{\thesection.\arabic{subsection}}

\makeatletter
\renewcommand{\p@subsection}{}
\renewcommand{\p@subsubsection}{}
\let\@makefntextOrig\@makefntext
\def\@makefntext#1{\@makefntextOrig{%
\baselineskip=15pt
#1}}
\makeatother

\usepackage{enumitem}
\usepackage[pdftex]{graphicx}
\usepackage{bm}
\usepackage[pdftex]{hyperref}
\usepackage[svgnames]{xcolor}
\definecolor{phthaloblue}{rgb}{0.0, 0.06, 0.54}
\hypersetup{
    colorlinks=true,
    linkcolor=phthaloblue,
    citecolor=blue,
    filecolor=blue,
    urlcolor=phthaloblue,
    }

\newcommand{\beq}{\begin{eqnarray}} 
\newcommand{\eeq}{\end{eqnarray}}

\newcommand{\bel}[1] {\begin{equation}\label{#1}}
\newcommand{\beal}[1] {\begin{eqnarray}\label{#1}}
\newcommand{\be}{\begin{equation}}
\newcommand{\ee}{\end{equation}}
\newcommand{\bea}{\begin{array}} 
\newcommand{\eea}{\end{array}}

\newcommand{\vev}[1]{ \left\langle {#1} \right\rangle }

\newcommand{\eV}{ \ {\rm eV} }
\newcommand{\keV}{ \ {\rm keV} }
\newcommand{\MeV}{\  {\rm MeV} }
\newcommand{\GeV}{\  {\rm GeV} }
\newcommand{\TeV}{\  {\rm TeV} }

\begin{document}

\title{Gravitational Waves and Dark Photon Dark Matter \\ from Axion Rotations}

\author{Raymond T. Co}
\affiliation{Leinweber Center for Theoretical Physics, Department of Physics, University of Michigan, Ann Arbor, MI 48109, USA}
\affiliation{William I. Fine Theoretical Physics Institute, School of Physics and Astronomy, University of Minnesota, Minneapolis, MN 55455, USA}
\author{Keisuke Harigaya}
\affiliation{School of Natural Sciences, Institute for Advanced Study, Princeton, NJ 08540, USA}
\author{Aaron Pierce}
\affiliation{Leinweber Center for Theoretical Physics, Department of Physics, University of Michigan, Ann Arbor, MI 48109, USA}

\date{\today}

\begin{abstract}
 An axion rotating in field space can produce dark photons in the early universe via tachyonic instability. This explosive particle production creates a background of stochastic gravitational waves that may be visible at pulsar timing arrays or other gravitational wave detectors. This scenario provides a novel history for dark photon dark matter. The dark photons may be warm at a level detectable in future 21-cm line surveys. For a consistent cosmology, the radial direction of the complex field containing the axion must be thermalized. We explore a concrete thermalization mechanism in detail and also demonstrate how this setup can be responsible for the generation of the observed baryon asymmetry. 
\end{abstract}

\preprint{LCTP-21-08, UMN-TH-4010/21, FTPI-MINN-21-03}

\maketitle

\begingroup
\hypersetup{linkcolor=black}
\renewcommand{\baselinestretch}{1.12}\normalsize
\tableofcontents
\renewcommand{\baselinestretch}{2}\normalsize
\endgroup

\newpage

\section{Introduction}
\label{sec:intro}

Spontaneously broken global symmetries are essential ingredients in solutions to a variety of problems in the Standard Model.  Examples include a Peccei-Quinn symmetry~\cite{Peccei:1977hh} (a solution to the strong CP problem), lepton symmetry~\cite{Chikashige:1980ui} (the origin of neutrino masses), and flavor symmetries~\cite{Froggatt:1978nt} (the  origin of the pattern of the fermion masses and mixing).  Associated with the breaking of such symmetries are light degrees of freedom--Nambu-Goldstone bosons.  The resulting particles go by various names: a QCD axion~\cite{Weinberg:1977ma,Wilczek:1977pj} for the Peccei-Quinn symmetry, a Majoron for lepton number~\cite{Chikashige:1980ui}, and a familon~\cite{Reiss:1982sq,Wilczek:1982rv} in the case of a flavor symmetry. Here we refer to these Nambu-Goldstone bosons generically as ``axions." 

It is commonly assumed that an axion field is initially static and begins oscillation when its mass becomes comparable to the Hubble expansion rate. However, the axion field may instead be initially rotating in field space, with important consequences. Such rotation may be initiated by an explicit breaking of the global symmetry coming from a higher dimensional operator containing the symmetry breaking field.  When the radial direction of the symmetry breaking field takes on large values, these operators are of increased importance and can give the axion a kick.  These ingredients are generic and are found, e.g., in Affleck-Dine baryogenesis~\cite{Affleck:1984fy}.  This setup has been studied in the context of baryogenesis: axiogenesis~\cite{Co:2019wyp}, ALP-genesis\cite{Co:2020xlh}, and lepto-axiogenesis~\cite{Co:2020jtv}.  Its impact on the axion dark matter abundance has been explored under the name of the kinetic misalignment mechanism~\cite{Co:2019jts,Co:2020dya}. In the above examples, axions directly couple to Standard Model particles.

However, if the coupling of the axion to particles in our sector is too weak, it can prove challenging to probe.  In this case, one may have to rely on signals from gravitational forces, e.g., superradiance~\cite{Brito:2015oca} and mini-clusters~\cite{Hogan:1988mp}, or perhaps signals from dark sectors. Here, we consider a rotating axion that couples to a dark photon and explore two consequences: a gravitational wave (GW) signal and the production of dark photon dark matter.

When the axion undergoes motion, it will modify the dispersion relation of a dark photon to which it couples.  If the coupling is sufficiently large, the dark photons become tachyonic for a range of dark photon wave numbers. In the conventional case where the axion is assumed to oscillate, the production of dark photons via tachyonic instability is ineffective for natural coupling strengths. Nevertheless, if the coupling is much larger, dark photons are explosively produced~\cite{Agrawal:2017eqm}. The produced dark photons are highly inhomogeneous and source gravitational waves~\cite{Machado:2018nqk, Machado:2019xuc}. This scenario was recently considered to explain the North American Nanohertz Observatory for Gravitational Waves (NANOGrav) signal~\cite{Arzoumanian:2020vkk}, but it was found that the axion oscillation leads to an excessive amount of dark matter~\cite{Kitajima:2020rpm} unless the axion mass decreases after the beginning of the oscillations~\cite{Namba:2020kij,Ratzinger:2020oct}. The produced dark photons may be dark matter~\cite{Agrawal:2018vin, Bastero-Gil:2018uel, Co:2018lka}.

As we will explore in this work, an axion that rotates in field space can also produce dark photons via tachyonic instability.  In contrast to the oscillation case, the sign of the velocity of the axion field does not alternate.  A consequence is that dark photon production may occur without any need to enhance the axion-dark photon coupling.
The tachyonic instability from a rotating field was invoked as way to produce the hypercharge gauge field in the context of magnetogenesis~\cite{Kamada:2019uxp}.
Furthermore, the rotational velocity, which determines the strength of the GW signal, is not directly tied to the axion mass. In this case, the axion may be nearly massless around the minimum of the potential, so the NANOGrav potential signal can be explained without overproduction of axion dark matter.  Throughout this work, we will assume such a nearly massless axion, although the mechanism may also work for a massive axion such as the QCD axion. This mechanism can produce GW signals in a wide range of frequencies of interest to current and future GW detectors such as NANOGrav~\cite{Arzoumanian:2018saf}, the Square Kilometer Array (SKA)~\cite{Janssen:2014dka}, the Laser Interferometer Space Antenna (LISA)~\cite{Audley:2017drz}, the DECi-hertz Interferometer Gravitational wave Observatory (DECIGO)~\cite{Sato:2017dkf}, and the Big Bang Observatory (BBO) \cite{Crowder:2005nr}. We study under what assumptions the dark photon can comprise the dark matter and when an alternate candidate is needed.  In the latter case, the dark photon comprises dark radiation.

As mentioned above, the axion field's rotation may be induced by the combination of a large field value in the radial direction and an explicit breaking of the symmetry.  In this case, radial motion is also expected. To avoid cosmological difficulties such as a moduli problem, radial motion should be dissipated via thermalization.  The necessity for thermalization places non-trivial constraints on the theory.

In the concrete thermalization model we consider, couplings between the axion and Standard Model fermions are generated. The global charge asymmetry in the axion rotation is partially translated into the particle-antiparticle asymmetry in the thermal bath, including the baryon asymmetry~\cite{Co:2019wyp,Co:2020xlh}. We show that the observed baryon asymmetry can be simultaneously explained in some of the parameter space.

The structure of the paper is as follows. In Sec.~\ref{sec:axion_dynamics}, we discuss the dynamics of the axion, including its rotation and how its coupling to a dark photon can lead to tachyonic instability.  
In Sec.~\ref{sec:GWrotations}, we discuss how the strength and frequency of produced GWs are related to the underlying axion parameters.  In Sec.~\ref{sec:DPDM}, we turn to a discussion of when and how the dark photon may comprise the dark matter in this model. 
The warmness of the dark photon gives an important constraint as well as a possible future signal.
In Sec.~\ref{sec:model}, we investigate a specific realization of this scenario, paying particular attention to the thermalization of the radial component of the symmetry breaking field and the baryon asymmetry produced from the axion rotation.  Technical details on the cosmology are postponed to Appendix~\ref{app:concrete_cosmo}. Sec.~\ref{sec:discussion} is devoted to the summary of the results and discussion.

\section{Dynamics of Axions and Dark Photons}
\label{sec:axion_dynamics}

\subsection{Axion rotations}
\label{sec:rotations}

We consider a field theoretical axion, a pseudo-Nambu-Goldstone boson associated with the spontaneous breaking of a global $U(1)_P$ symmetry.
The axion $\phi$ is the angular direction of a complex scalar field $P$,
\begin{align}
    P = \frac{1}{\sqrt{2}} \, S \,  e^{i \theta},
\end{align}
where $S$ is the radial direction that we refer to as the saxion. Its value at the minimum of the potential determines the decay constant $f_\phi=\vev{S}$, and $\theta = \phi/f_\phi$ is the angular direction.

If, as conventionally assumed, the axion field has zero initial velocity, it begins oscillations when its mass is comparable to the Hubble expansion rate, $m_{\phi} \sim H$. This picture, however, may be too simplistic. In the early universe, the saxion field value may be much larger than $f_{\phi}$. For instance, the saxion may be dynamically driven to a large field value due to a negative Hubble-induced mass during inflation~\cite{Dine:1995uk}.  The field value may be stabilized by a higher dimensional operator. When the saxion takes on such large field values, it is not guaranteed that the potential of $P$ remains nearly $U(1)_P$ symmetric. A higher dimensional potential term that explicitly breaks the global symmetry,
\begin{align}
    \Delta V \supset \frac{P^{n}}{M^{n-4}} + {\rm h.c.},
\end{align}
where $M$ is a cut-off scale, is enhanced at large field values.  Such terms are in fact expected when the $U(1)_P$ symmetry accidentally arises as a result of another exact symmetry~\cite{Holman:1992us,Barr:1992qq,Kamionkowski:1992mf,Dine:1992vx}. This explicit breaking can initiate an angular motion for $P$. Cosmic expansion will cause the saxion field value to decrease, with an accompanying improvement in the quality of the global symmetry. Once the global symmetry is approximately conserved, the field $P$ continues to rotate so as to preserve the angular momentum in field space.  This is because the angular momentum may be identified with the charge of the global symmetry associated with the rotation, $n_\theta$.
The rotation of a complex field induced by a higher dimensional operator was originally considered in the context of Affleck-Dine baryogenesis~\cite{Affleck:1984fy}.
As we will discuss below, this rotation has important cosmological consequences, including the possible production of gravitational waves and dark matter.

In general, the initial rotation is not perfectly circular--it contains both angular and radial motion.  If the radial motion is not damped, it can come to dominate the energy density of the universe.  Here we assume that $P$ couples to the thermal bath in a way that eventually allows for its thermalization and the removal of this possible moduli problem. We will discuss an  explicit realization of this thermalization mechanism in Sec.~\ref{sec:model}.
Upon thermalization, the rotation becomes circular (up to thermal fluctuations). Even if the thermalization occurs via scattering with $U(1)_P$-charged particles so that the rotation could lose its charge, most of the charge is maintained in the rotation, since that is the state with the least free energy~\cite{Co:2019wyp}. In this paper, we consider the scenario where thermalization occurs sufficiently early so that the motion is circular during the era of tachyonic instability.

The circular motion then evolves as follows~\cite{Co:2019wyp}. The angular velocity is determined by the equation of motion. Long after the  beginning of the rotations, when $V'(S)/S \gg H^2$ with $V(S)$ the potential energy of $S$,
\begin{align}
\label{eq:theta_evolve}
 \dot{\theta}^2 = \frac{V'(S)}{S}.
\end{align}
The evolution of $S$ and the equation of state of the rotation may be determined from charge conservation, 
\begin{align}
    n_\theta = \dot{\theta} S^2 =  \sqrt{V'(S)}S^{3/2}  \propto R^{-3},
\end{align}
where $R$ is the scale factor of the universe. For example, if $V(S)$ is quartic and $S\gg f_\phi$, 
\begin{equation}
\label{eq:scaling_radiation}
    S\propto R^{-1}, \qquad \dot{\theta} \propto R^{-1},   \qquad \rho_\theta \propto R^{-4}, \qquad   \qquad   {\rm for}~V(S) \propto S^4,
\end{equation}
where $\rho_\theta \equiv \dot\theta^2 S^2$ is the energy density associated with the rotation. If $V(S)$ is nearly quadratic and $S\gg f_\phi$, 
\begin{equation}
\label{eq:scaling_matter}
    S\propto R^{-3/2}, \qquad  \dot{\theta} \propto R^{0}, \qquad \rho_\theta \propto R^{-3}, \qquad  \qquad    {\rm for}~V(S) \propto S^2.
\end{equation}
Once the saxion reaches its minimum, $S\simeq f_\phi$, 
charge conservation requires 
\begin{equation}
\label{eq:scaling_kination}
    S\propto R^{0}, \qquad  \dot{\theta} \propto R^{-3}, \qquad \rho_\theta \propto R^{-6}, \qquad  \qquad     {\rm for}~S\simeq f_\phi.
\end{equation}
Then the energy density of the rotation $\rho_\theta$ scales as so-called kination~\cite{Spokoiny:1993kt,Joyce:1996cp}. 

In the following subsection, we discuss the consequences of axion rotation when the axion couples to a dark gauge field.

\subsection{Tachyonic instability}
\label{sec:TI}

\subsubsection{Dark photon production}
If the global symmetry has a mixed quantum anomaly with a gauge interaction, the axion couples to the corresponding gauge field. We consider the case where an anomaly is present for a dark $U(1)_D$ gauge interaction, resulting in a coupling

\begin{equation}
    \label{eq:axion-DP}
    {\mathcal L } \supset \frac{e_D^2}{64\pi^2} \theta \epsilon^{\mu\nu \rho \sigma} F'_{\mu\nu}F'_{\rho \sigma},
\end{equation}
where $e_D$ is the dark gauge coupling and $F'$ is the field strength of $U(1)_D$. The mixed quantum anomaly arises from the coupling of $P$ to fermions $\psi \bar{\psi}$ simultaneously charged under $U(1)_D$ and global $U(1)_P$,
\begin{equation}
\label{eq:mediator}
    {\mathcal L} \supset y_\psi P \bar{\psi} \psi + {\rm h.c.} .
\end{equation}

The equations of motion for the two transverse modes of the dark photon, $A_{\pm}'$, are
\begin{align}
    \frac{\partial^2 A_{\pm}' }{\partial t^2} + H \frac{\partial A_{\pm}' }{\partial t} + \left( m_{A'}^2 + \frac{k^2}{R^2} \pm \frac{e_D^2  }{8 \pi^2} \frac{k}{R} \dot{\theta}  \right) A_{\pm}' = 0 \,, 
\end{align}
where $m_{A'}$ is a possible dark photon mass and $H$ is the Hubble expansion rate.
The dark photon has a tachyonic instability centered around the momentum 
\begin{align}
\label{eq:thetadot_k}
  \frac{k}{R} \simeq \frac{e_D^2}{16\pi^2} \dot{\theta} \equiv k_{\rm TI},
\end{align}
if $m_{A'} < k_{\rm TI}$. Inside the instability band, the growth rate of the dark photon fluctuation is as large as $k_{\rm TI}$.
The global charge $n_\theta$ is transferred to the dark photon helicity density.

If the axion field has negligible initial velocity in field space, it begins oscillating when $H \sim m_\phi$.  The sign of $\dot{\theta}$ then alternates with a frequency $m_\phi$, so efficient tachyonic instability requires $k_{\rm TI} > m_\phi \sim \dot{\theta}$.
By the above equation, it can be seen that unless $e_D \sim 4\pi$, the tachyonic instability is ineffective. This conclusion may be avoided if the axion-dark photon coupling is much larger than Eq.~(\ref{eq:axion-DP}), so that $k_{\rm TI}$ is comparable to $m_\phi$ at the beginning of the oscillations. 
Such an enhancement is possible by utilizing the Kim-Nilles-Peloso mechanism~\cite{Kim:2004rp} and its  clockwork generalization~\cite{Harigaya:2014eta,Choi:2014rja,Harigaya:2014rga,Choi:2015fiu,Kaplan:2015fuy}, as considered in~\cite{Farina:2016tgd}.

On the other hand, if the axion field is initially rotating, the sign of $\dot{\theta}$ remains the same. Also, for $S>f_\phi$, the angular velocity evolves as in Eq.~(\ref{eq:theta_evolve}), and it may decrease more slowly than the Hubble expansion rate. As a result, the growth rate via tachyonic instability, as set by the $k_{\rm TI}$ of Eq.~(\ref{eq:thetadot_k}), may eventually outpace the Hubble expansion.  The consequence is explosive dark photon production--without resorting to an enhanced axion-dark photon coupling via baroque model building.
Note that once $S=f_\phi$, $\dot{\theta}$ decreases in proportion to $R^{-3}$ as in Eq.~(\ref{eq:scaling_kination}), and the tachyonic instability will not become effective if it has not already done so. 

Once $k_{\rm TI} >H$, the dark photon density is exponentially amplified, seeded by quantum fluctuations.
This continues until back-reactions stop the tachyonic instability.
This can occur either when the angular velocity decreases substantially via the loss of $n_\theta$ stored in $P$ or when the production of axion fluctuations from dark photons changes the angular velocity.
The former can occur when an ${\mathcal O}(1)$ fraction of the energy of the rotation is transferred into the dark photon. Computation of the scattering rate between $A'$ and the rotating field, taking care to include Bose enhancement for $A'$, shows that the latter is relevant at a similar time.
We assume that the production of dark photons from the axion rotation is saturated when ${\mathcal O(1)}$ of the axion energy/number density is transferred, leaving a rigorous examination by lattice computation for future work. We note that if $S$ is already close to $f_\phi$ when the ${\mathcal O}(1)$ transfer occurs, the tachyonic instability necessarily rapidly becomes ineffective since $\dot{\theta}$ then decreases faster than $H$ does according to Eq.~(\ref{eq:scaling_kination}).
The Hubble scale when the transfer occurs is
\begin{equation}\label{eq:kTIvsHubble}
 H_p = \frac{k_{\rm TI}}{r_p} ,
\end{equation}
where $r_p$ is an $\mathcal{O}(10)$ factor that allows for the exponential amplification. The subscript $p$ indicates evaluation at the time of dark photon production so that $S_p$, $\dot{\theta}_p$, and $T_p$ denote the saxion field value, the angular velocity, and temperature at production. Determining the precise spectrum of the dark photons requires a numerical lattice simulation beyond the scope of this paper. Motivated by the detailed study of tachyonic instability in other contexts~\cite{Kitajima:2020rpm,Ratzinger:2020oct}, we assume that when $H=H_p$, the momentum distribution of dark photons is sharply peaked at $k_{\rm TI}$ and the dark photon energy density is as large as the  energy stored in the rotations at the onset of the tachyonic instability, $\rho_\theta$. 

\subsubsection{Axion rotations and fluctuations}

We now discuss the fate of the axion field after the dark photon production. In this paper, we assume that the axion is massless but comment on the possibility of a massive axion and axion dark matter in Sec.~\ref{sec:discussion}.

Even after the production of dark photons via tachyonic instability is saturated,  non-zero angular momentum $n_\theta$ remains in the axion rotation.
The rotation should not dominate the energy density of the universe during Big-Bang Nucleosynthesis (BBN) or after recombination. As we will see in Sec.~\ref{sec:model}, this can impose a non-trivial constraint.
We parameterize the residual rotation by
\begin{align}
\label{eq:rtheta}
 Y_{\theta} = \frac{n_{\theta}}{s} = r_\theta Y_{\theta,i},
\end{align}
where $Y_{\theta,i}$ is the yield of the global charge before tachyonic instability becomes effective. We expect that $r_{\theta} ={\mathcal O} (1)$ if the dark photon production is saturated when $S$ is close to $f_\phi$, since as argued above Eq.~(\ref{eq:kTIvsHubble}), tachyonic instability rapidly becomes ineffective after $S$ reaches its minimum.  
Even for $S > f_\phi$ at the time of production, it is plausible that after the saturation of the production of dark photons, the tachyonic instability ceases when the scattering between axions and dark photons makes the rotation incoherent so that $r_\theta = \mathcal{O}(1)$.  We take $r_\theta$ as a free parameter, leaving the rigorous determination of $r_\theta$ by a numerical simulation to future work.

The produced dark photons can create axion fluctuations via the coupling in Eq.~(\ref{eq:axion-DP}).  As long as the axions are light, as assumed here, the axions produced in this way at most have the same energy as dark photons and contribute to dark radiation, which will be discussed around Eq.~(\ref{eq:DeltaNeff}).

\subsubsection{Effects of $U(1)_{D}$-charged fields} 
In the above discussion, we have integrated out the $U(1)_D$-charged fermion $\psi$ and ignored its dynamics. However, it could conceivably be produced in several ways, as enumerated below. Should this occur prior to the era of tachyonic instability, electric conductivity from the fermions~\cite{Baym:1997gq,Arnold:2000dr,Domcke:2018eki} acts as a friction term in the equation of motion for the gauge field and may prevent efficient tachyonic instability from occurring. It is important to ensure none of these production modes are present.

\begin{enumerate}
\item 
The fermions can be produced from scattering between the particles in the thermal bath. This can be avoided if either $y_\psi S \gg T$ so that the production of $\psi$ is kinematically forbidden, or the couplings of $\psi$ with the bath are sufficiently small that the production is inefficient.  
\item
When the motion of $P$ is not completely circular, the mass of $\psi$ oscillates and $\psi$ may be produced by parametric resonance. However, as long as $y_\psi S \gg \dot{\theta}$, the oscillation of the mass is adiabatic, and production by parametric resonance is exponentially suppressed.
\item
The fermions could be produced by the Schwinger effect from the dark electric field $E_D$~\cite{Heisenberg:1935qt,Schwinger:1951nm}, or by the chiral anomaly from the dark helical electromagnetic field~\cite{Adler:1969gk,Bell:1969ts,Domcke:2018eki}.
This production is exponentially suppressed if the fermion mass $y_\psi S$ is larger than $(e_D E_D)^{1/2}$ at the time of gauge field production.
\end{enumerate}
In the concrete realization discussed in Sec.~\ref{sec:model}, we impose the conditions listed above but find that only those in 1) give relevant constraints because $T\gg \dot{\theta}, E_D^{1/2}$. When we identify the produced dark photons as dark matter, we also impose these conditions even after the dark photons are produced.  This ensures that energy is not removed from the dark photons by electric conductivity or via the fermion production discussed in 3).

When the condition in 3) is satisfied, this also ensures that the mass of $\psi$ is larger than $k_{\rm TI}$,  justifying integrating out the fermion.  It is therefore consistent to use the dimension-five operator in Eq.~(\ref{eq:axion-DP}).
After $\psi$ is integrated out, the dark gauge field interacts via a dimension-eight operator suppressed by the mass of $\psi$ called the Euler-Heisenberg term~\cite{Heisenberg:1935qt}. It is natural to ask whether these self-interactions cause any difficulties.  However, the energy density from the Euler-Heisenberg term is much smaller than the energy density from the kinetic term as long as $y_\psi S <(e_D E_D)^{1/2}$.  Therefore, it is not expected that this term causes any difficulty for efficient tachyonic instability. Furthermore, while self-scattering via the Euler-Heisenberg term may affect the spectrum of the dark photons, in the concrete realization we discuss in Sec.~\ref{sec:model}, we find that the scattering rate is smaller than the Hubble expansion rate, so it is irrelevant.

\subsubsection{Deviation from coherent circular motion}\label{sec:thermalizationComm}

In the above analysis, we have assumed that the rotation is coherent and circular. Indeed, in this paper, we require that thermalization of the rotation occur at a temperature above $T_p$ so that the motion is nearly circular by the time of tachyonic instability. 

First, we comment on a potential subtlety. After the beginning of the rotation and before the completion of thermalization, the rotation is not yet circular. During this phase, it is possible that parametric resonance~\cite{Dolgov:1989us, Traschen:1990sw, Kofman:1994rk, Shtanov:1994ce, Kofman:1997yn} may produce fluctuations of $P$ with wave numbers $\sim \sqrt{V''(S)}$ ~\cite{Co:2020dya,Co:2020jtv}. If there is an epoch where parametric resonance becomes efficient before the completion of thermalization, the motion of $P$ is no longer coherent and has fluctuating components with a field magnitude $\sim S$. However, thanks to charge conservation, the field $P$ still rotates as a whole. Once thermalization occurs, the fluctuations are dissipated, and the field value of $P$ is nearly homogeneous within a horizon (up to thermal fluctuations) and continues to rotate. Tachyonic instability will then proceed as described above.

We now make a few brief comments on what might happen in the case where thermalization has not occurred by the era of tachyonic instability.  As we now discuss, it is plausible that tachyonic instability still occurs, which might allow a relaxation of our requirement that thermalization occur prior to tachyonic instability.   Of course, thermalization would in any case ultimately be necessary to avoid a moduli problem.  First we consider the case that parametric resonance has not occurred before the epoch of tachyonic instability.  In this case, the rotation is coherent,  but owing
to non-thermalized motion in the radial direction, $\dot{\theta}$ oscillates with a frequency $\sim \sqrt{V''(S)}$. Despite this, modes with $k \ll \sqrt{V''(S)}$ only feel the time-averaged angular velocity $\vev{\dot{\theta}} \sim \sqrt{V''(S)}$ so that tachyonic instability still occurs. We have confirmed this by numerically solving the equation of motion of the dark photon. 

We next consider the case where the parametric resonance becomes effective before the time of tachyonic instability, but thermalization has not yet occurred.
Although the field value of $P$ fluctuates, we expect that the dark photons with wave numbers $\ll \sqrt{V''(S)}$ feel the spatial- and time-averaged $\dot{\theta}$, which is ${\mathcal O}(\epsilon_{r} \sqrt{V''(S)})$ because of charge conservation, where $\epsilon_{r}$ is the ratio between the $U(1)_P$ charge density and the number density of the radial direction. We expect that tachyonic instability still occurs. 

Despite these plausibility arguments for an effective tachyonic instability even in the presence of delayed thermalization, for simplicity, in our analysis, we will require thermalization to occur before the dark photon production by tachyonic instability.

\section{Gravitational Waves from Rotations} \label{sec:GWrotations}

Gravitational waves are created at the epoch of explosive vector field production \cite{Machado:2018nqk, Namba:2020kij, Kitajima:2020rpm}.  The GW equation of motion is given by
\begin{equation}\label{gravityEOM}
\ddot{h}_{ij} + 3 H \dot{h}_{ij} - \frac{1}{a^2} \nabla^2 h_{ij} = \frac{2}{M_{\rm Pl}^2} \Pi_{ij}^{TT},
\end{equation}
where $M_{\rm Pl}=2.4 \times 10^{18} \GeV$ and $\Pi_{ij}^{TT}$ represents the transverse-traceless anisotropic stress tensor.
Importantly, $\Pi_{ij}^{TT}$  receives a contribution from the quadrupole of the energy of the dark photon, which sources the GWs.

The peak frequency of the GW $f_{\rm GW}$ is set by that of the dark photons at the time of creation. In Eq.~(\ref{eq:kTIvsHubble}), we defined $r_p$ as the constant of proportionality that relates the dark photon wave number to the horizon.  We expect that the GW wave number at emission is given by $k_{\rm GW} \approx k_{\rm TI} \equiv r_p H_p$. This redshifts to the present day, resulting in a peak frequency
\begin{equation}
\label{eq:fGW_today}
f_{\rm GW}  = 2~{\rm nHz} \; \left( \frac{r_p}{20} \right)^{ \scalebox{1.01}{$\frac{1}{2}$} } \left(\frac{k_{\rm TI}}{3 \times 10^{-13}  \, \rm{eV}} \right)^{ \scalebox{1.01}{$\frac{1}{2}$} } \left(\frac{10}{g_*(T_p)}\right)^{ \scalebox{1.01}{$\frac{1}{12}$} }.
\end{equation}
Here, we have normalized to a frequency relevant for a signal at NANOGrav~\cite{Arzoumanian:2020vkk} or SKA~\cite{Janssen:2014dka}.  
The GW frequency determines the temperature at production
\begin{equation}
    \label{eq:Tprod}
    T_{p} \simeq 5.5 \MeV  \left(\frac{f_{\rm GW}}{\rm{2~nHz}}\right) \left( \frac{20}{r_p} \right) \left(\frac{10}{g_*(T_p)}\right)^{ \scalebox{1.01}{$\frac{1}{6}$} } .
\end{equation}

Gravitational waves are initially produced with energy density $\rho_{\rm GW}= M_{\rm Pl}^2 \langle \dot{h}_{ij} \dot{h}_{ij} \rangle/4$.  This energy density redshifts as that of radiation $\rho_R$; the ratio $\rho_{\rm GW}/\rho_R$ is invariant.  Examination of Eq.~(\ref{gravityEOM}) allows us to evaluate this quantity at the time of production as
\begin{equation}
\label{eq:GWoverRad}
    \frac{\rho_{\rm GW}}{\rho_R}= \frac{x^{2} \rho_{A^{\prime}}^2}{3 k_{\rm TI}^4 M_{\rm Pl}^4}=\frac{x^{2} \dot{\theta}_p^4 S_{p}^4}{3 k_{\rm TI}^4 M_{\rm Pl}^4},
\end{equation}
with $k_{\rm TI}$ the wave number at production as defined above.  In the second equality, we have noted that the energy density in dark photons at production $\rho_{A^{\prime}}$ is expected to be of the same order of magnitude as the energy in the rotation prior to the tachyonic instability, $\rho_\theta \sim \dot{\theta}_p^2 S_{p}^2$.  Here the subscript on $\dot{\theta}_p$ emphasizes that this quantity should be evaluated at the time of the tachyonic instability production.  
This results in a present day GW energy density at the peak frequency $f_{\rm GW}$
\begin{align}
\label{eq:GWSignal}
\Omega_{\rm GW} h^2 & \simeq 3 \times 10^{-9} \left( \frac{20}{r_p} \right)^2 \left(\frac{10}{g_*(T_p)}\right)^{ \scalebox{1.01}{$\frac{1}{3}$} } \left. \left( \frac{\rho_\theta}{0.1\rho_R} \right)^2 \right|_{H = H_p} \\ \nonumber
& \simeq 3 \times 10^{-10} \left(\frac{\rm{2~nHz}}{f_{\rm GW}}\right)^8 \left(\frac{\dot{\theta}_p}{\rm MeV}\right)^4 \left( \frac{S_p}{10 \MeV}\right)^4  \left( \frac{r_p}{20} \right)^6 \left(\frac{10}{g_*(T_p)}\right) \\ \nonumber
& \simeq 4 \times 10^{-13} \left(\frac{0.1~\rm{Hz}}{f_{\rm GW}}\right)^8 \left(\frac{\dot{\theta}_p}{100 \GeV}\right)^4 \left( \frac{S_p}{10^8 \GeV}\right)^4  \left( \frac{r_p}{20} \right)^6 \left(\frac{200}{g_*(T_p)}\right).
\end{align}
In our explicit realization in Sec.~\ref{sec:model}, we show a case in which $\dot{\theta}_p \simeq m_S$ with $m_S$ the saxion mass. Here in the second (third) line, we have normalized the equation to a possible signal at NANOGrav (and at DECIGO/BBO), but this scenario can instead give rise to signals visible at other pulsar timing arrays such as SKA or space-based GW detectors such as LISA.  To observe a signal in a similar frequency range but with a smaller amplitude than that suggested by NANOGrav, it may be necessary to efficiently subtract ``background" events from super massive black hole mergers~\cite{Breitbach:2018ddu,Regimbau:2016ike}.  

For the dark photon spectrum sharply peaked at $k_{\rm TI}$, the GW spectrum is peaked around the frequency $f_{\rm GW}$ in Eq.~(\ref{eq:fGW_today}) with a magnitude in Eq.~(\ref{eq:GWSignal}).  It decreases in proportion to $f^{3}$ for frequency $f < f_{\rm GW}$~\cite{Caprini:2007xq}, and is sharply cut off above $f_{\rm GW}$. The scattering of the dark photon and the axion will make the spectrum broader~\cite{Kitajima:2020rpm,Ratzinger:2020oct}. The scattering is expected to be more efficient than the case of axion oscillations as we will discuss in Sec.~\ref{sec:DPDM}.
Also, the polarization of the GWs could differ from the oscillation case. Here, a linear combination of the $U(1)_P$ charge in the axion rotation and the helicity of the dark photons is conserved, so strong polarization is expected to persist even in the presence of scattering and backreaction. 
Finally, should non-negligible  production of dark photons from the axion rotation occur even after the transfer of ${\mathcal O}(1)$ fraction of the axion energy, contrary to our assumption, the spectrum will become broadened towards higher frequencies.

From the relationship between $k_{\rm TI}$ and the angular velocity given in Eq.~(\ref{eq:thetadot_k}), the dependence of the dark gauge coupling on the GW signal is as follows:
\begin{align}
\label{eq:gaugecoupling}
e_D 
&\simeq 6 \times 10^{-9} \left( \frac{f_{\rm GW}}{2~{\rm nHz}} \right) \left( \frac{\rm MeV}{\dot{\theta}_p} \right)^{ \scalebox{1.01}{$\frac{1}{2}$} } \left( \frac{20}{r_p} \right)^{ \scalebox{1.01}{$\frac{1}{2}$} } \left(\frac{g_*(T_p)}{10} \right)^{ \scalebox{1.01}{$\frac{1}{12}$} } \\
& \simeq 10^{-3} \left( \frac{f_{\rm GW}}{0.1~{\rm Hz}} \right) \left( \frac{100 \GeV}{\dot{\theta}_p} \right)^{ \scalebox{1.01}{$\frac{1}{2}$} } \left( \frac{20}{r_p} \right)^{ \scalebox{1.01}{$\frac{1}{2}$} } \left(\frac{g_*(T_p)}{200} \right)^{ \scalebox{1.01}{$\frac{1}{12}$} }. \nonumber
\end{align}
For a fixed magnitude of GWs and smaller $\dot{\theta}_p$, the field value $S_p$ is required to be larger according to Eq.~(\ref{eq:GWoverRad}). The coupling of $P$ with the thermal bath then becomes weaker, and hence $\dot{\theta}_p$ is bounded from below for the thermalization of the radial mode to be successful. We will see this explicitly in Sec.~\ref{sec:model}. For frequencies of GWs relevant for pulsar timing arrays, a small dark gauge coupling constant is required.

Depending on the mass of the dark photon, it may act as either dark radiation or dark matter. The possibility of dark photon dark radiation from tachyonic instability is also discussed in Refs.~\cite{Agrawal:2017eqm,Machado:2018nqk,Ratzinger:2020oct}. In the (nearly) massless case, the contribution to $N_{\rm eff}$ is intimately tied to the size of the GW signal.  This  is because the abundance of GW is related to the dark photon production. 
We can write
\begin{equation} 
\label{eq:DeltaNeff}
\Delta N_{\rm eff} = \frac{4 r_{p} g_*(T_p)  }{7 \sqrt{3} } \sqrt{\frac{\rho_{\rm GW} }{\rho_R}}  \simeq 0.1  \left(\frac{\Omega_{\rm GW} h^{2} }{10^{-10}}\right)^{ \scalebox{1.01}{$\frac{1}{2}$} } \left( \frac{r_p}{20} \right) \left( \frac{g_*(T_p)}{10} \right)^{ \scalebox{1.01}{$\frac{7}{6}$} }.
\end{equation}
Axion fluctuations produced from the dark photons can also contribute to dark radiation, but their abundance does not exceed that in Eq.~(\ref{eq:DeltaNeff}).
Without a detailed lattice simulation dedicated to the case of an axion with rotation, such as the one done in the non-rotating case~\cite{Ratzinger:2020oct}, one cannot more precisely determine the relative energy transferred from the rotation to the GWs and the dark photon. For larger GW signals, a more precise computation could be of interest to establish a firm prediction for CMB-S4~\cite{Abazajian:2016yjj}.
 
\section{Dark Photon Dark Matter from Rotations}
\label{sec:DPDM}
If the dark photon is massive,%
\footnote{If the dark photon mass is generated via a Higgs mechanism, interactions with the Higgs boson must not spoil production via tachyonic instability.  Here we assume the charge of the dark Higgs is small enough that this is the case. We plan a full analysis of this constraint in a future work~\cite{CHP_DPDM}.   Even with such a small charge, the exchange of the dark Higgs generates a quartic coupling for the dark photon, which inhibits superradiance production of dark photons that could otherwise constrain light dark photons~\cite{Brito:2015oca}.}
it is an attractive dark matter candidate. Since the GWs are associated with dark photon production, the present day dark photon number density is related to the size of the GW signal.

It is important to check that dark photon dark matter does not thermalize with the Standard Model bath following its production via tachyonic instability.  Otherwise, the would-be dark matter abundance is depleted down to a thermal abundance, and while such a population would generally contribute an acceptably small contribution to $N_{\rm eff}$, it could not be cold dark matter.  This constraint depends on details of the implementation, including, e.g., the masses of the $U(1)_P$-charged fermions integrated out to generate the coupling between the axion and the dark photons.  In our specific realization in Sec.~\ref{sec:model}, we discuss this in more detail.  

The spectrum of dark photons are initially sharply peaked at the momentum $k_{\rm TI}$. The spectrum can, however, evolve via the scattering by the axion-dark photon coupling in Eq.~(\ref{eq:axion-DP}). The effect of the scattering is investigated with lattice simulations in Refs.~\cite{Kitajima:2020rpm,Ratzinger:2020oct} for the case of axion oscillations, and it is found that the dark photon momentum is approximately conserved up to the redshift. In our case, there are two important differences. First, there exist massless excitation modes around a background of a nearly circularly rotating $P$ even when $S>f_\phi$; these would correspond to axions at the minimum of the potential $S=f_\phi$.   In the oscillation case, the dark photon scattering rate at low momenta is suppressed by the axion mass, but no such suppression occurs here.
Second, the effective decay constant of the axion $\sim S$ continues to decrease after the production as shown by Eqs.~(\ref{eq:scaling_radiation}) and (\ref{eq:scaling_matter}) so that the scattering rate may increase, even after accounting for the dilution of the dark photon by Hubble expansion.  For these reasons, we expect that the scattering can be efficient and may change the spectrum appreciably.

The evolution of the spectrum of the dark photons may affect the predictions of the mass of the dark photons as well as their warmness. We first discuss the predictions neglecting the possible scattering as a baseline. We then qualitatively discuss the effects of the scattering and how the predictions may be altered, leaving a rigorous discussion via numerical simulations to future work. 

\subsection{The case without scattering}
\label{sec:DPDM_no_scatt}
We first estimate the warmness and mass of dark photons neglecting the scattering of dark photons following their production by tachyonic instability.
The amount of the GWs fixes the fractional energy density of dark photons before they become non-relativistic at $T=T_{\rm NR}$, after which the fractional energy density increases in proportion to $T^{-1}$.  Requiring dark photons to be dark matter, we can predict $T_{\rm NR}$ using Eq.~(\ref{eq:GWoverRad}),
\begin{equation}
\label{eq:DPDM_warmness}
T_{\rm NR} = \frac{4}{\sqrt{3}} \left(\frac{\rho_{\rm DM}}{s}\right) \left(\frac{g_{\ast}(T_{p})}{g_{\ast}(\rm{eV})}\right)^{ \scalebox{1.01}{$\frac{1}{3}$} } \left( \frac{\rho_R}{\rho_{\rm GW}}\right)^{ \scalebox{1.01}{$\frac{1}{2}$} }  \simeq 4 \keV \left(\frac{10^{-14}}{\Omega_{\rm GW} h^2} \right)^{ \scalebox{1.01}{$\frac{1}{2}$} } \left( \frac{20}{r_p} \right) \left(\frac{g_{\ast}(T_{p})}{10}\right)^{ \scalebox{1.01}{$\frac{1}{6}$} } ,
\end{equation}
with $\rho_{\rm DM}/s = 0.44$ eV the observed dark matter co-moving energy density.

Notably, the warmness exclusively depends on (the square root of) the magnitude of the GW signal. It is this quantity $\Omega_{\rm GW} h^2$ that sets the initial density in dark photon dark matter. If this initial density is larger (smaller), the dark photon must spend a smaller (larger) amount of time redshifting as matter to ensure that the right dark matter density $\rho_{\rm DM}/s$ will be obtained.  This, in turn, is directly related to the warmness of the dark matter. 

\begin{figure}
    \centering
    \includegraphics[width=0.8\columnwidth]{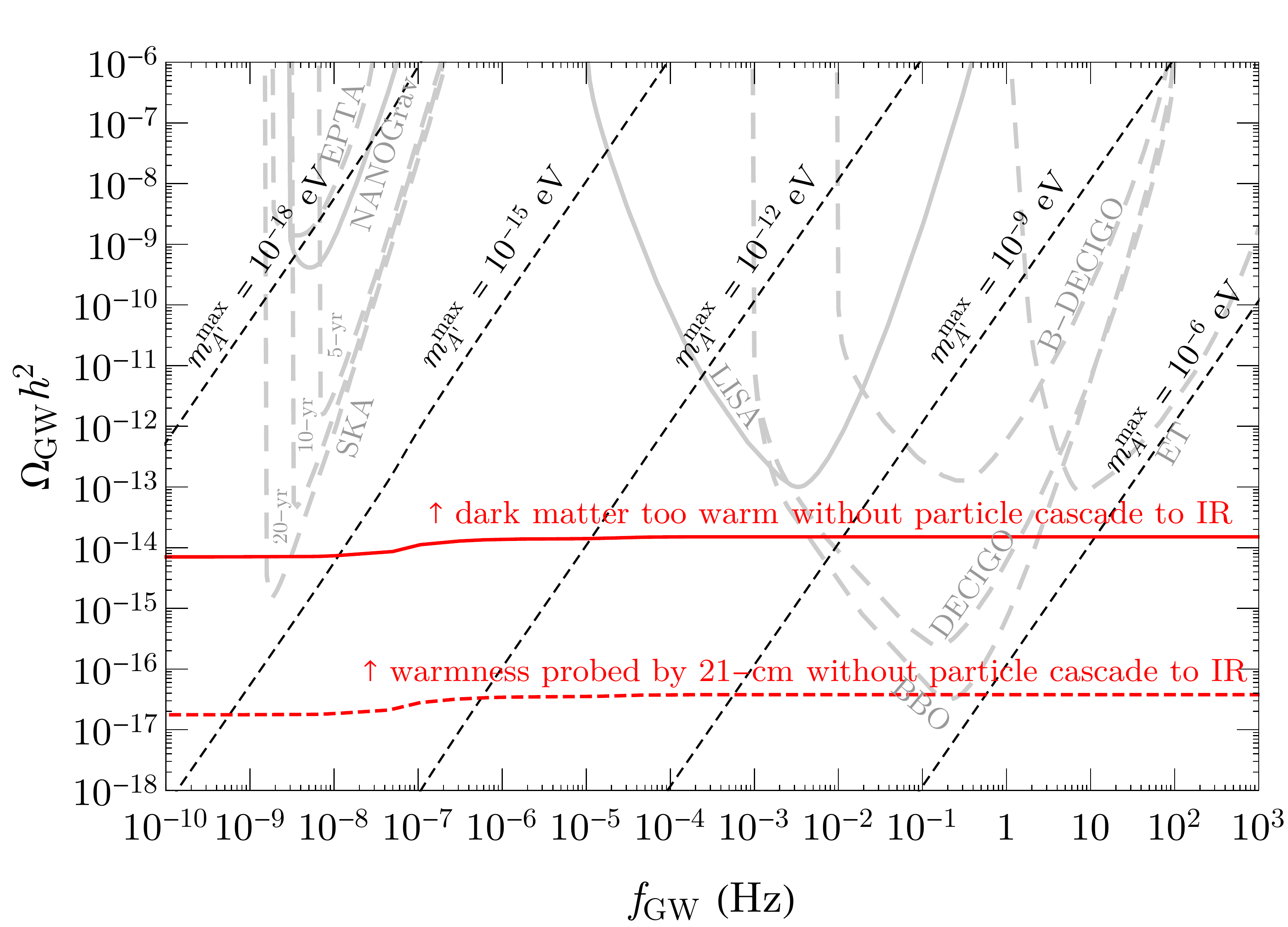}
    \caption{
     Correlations between GW signals and dark photon dark matter. In the limit that scatterings subsequent to production are neglected, the region above the red solid curve is excluded in the dark photon dark matter scenario because the produced dark photon would be too warm. Scatterings among dark photons and axions can relax this constraint--see Sec.~\ref{sec:DPDM_scatt}. Above the red dashed curve, the warmness can be probed via future 21-cm measurements~\cite{Sitwell:2013fpa}.
    The predictions of the dark photon mass are given by the black dashed lines.
    If scatterings increase the dark photon number, smaller dark photon masses will be predicted, and these contours should be regarded as the maximum possible values $m_{A'}^{\rm max}$. The Standard Model $g_*(T)$ is used, and the quantity $r_p$ defined in Eq.~(\ref{eq:kTIvsHubble}) is fixed at $r_p = 20$. The warmness constraint and the mass contours change with $r_p$ according to Eq.~(\ref{eq:DPDM_warmness}) and Eq.~(\ref{eq:DPDM_mass}). The sensitivity curves of the gravitational waves are provided by Ref.~\cite{Breitbach:2018ddu} for EPTA~\cite{Lentati:2015qwp}, NANOGrav~\cite{Arzoumanian:2018saf}, SKA~\cite{Janssen:2014dka}, LISA~\cite{Audley:2017drz}, (B-)DECIGO~\cite{Seto:2001qf, Sato:2017dkf}, BBO~\cite{Crowder:2005nr}, and ET~\cite{Sathyaprakash:2012jk}.
    }
    \label{fig:GW_DPDM}
\end{figure}

The current bound on the warmness arising from the Lyman-$\alpha$ measurements \cite{Irsic:2017ixq} indicates that the dark matter should become non-relativistic by a temperature
\begin{equation}
\label{eq:TNR_bound}
T_{\rm NR} > 5 \keV .
\end{equation}
Future measurements relying on observations of 21-cm radiation should be sensitive up to $T_{\rm NR} = 100 \keV $~\cite{Sitwell:2013fpa}.
This warmness constraint is shown in Fig.~\ref{fig:GW_DPDM}, where it shows up as the red solid curve--as expected from Eq.~(\ref{eq:DPDM_warmness}).  The small deviation from horizontal at low frequency is due to changes in $g_{\ast}$.   This figure indicates the dark photon dark matter scenario is inconsistent with a putative NANOGrav signal. 
However, a future GW signal from the SKA--or a satellite-based detector such as DECIGO--could be consistent with this framework.  Notably, a signal observed at SKA or DECIGO would result in dark photon dark matter with a finite residual velocity that could leave an imprint on structure formation that might be observed by future 21-cm line observations.  However, we emphasize that these conclusions may be modified by scattering in some cases, as discussed in the following subsection.

We obtain from Eqs.~(\ref{eq:fGW_today}) and (\ref{eq:GWoverRad}) the dark photon mass required to explain the observed dark matter abundance in the absence of scattering
\begin{equation}
\label{eq:DPDM_mass}
    m_{A^{\prime}} \simeq 10^{-8} \eV \left(\frac{10^{-14}}{\Omega_{\rm GW} h^{2} } \right)^{ \scalebox{1.01}{$\frac{1}{2}$} } \left(\frac{f_{\rm GW}}{0.1~{\rm Hz}} \right) \left( \frac{20}{r_p} \right) \left( \frac{g_*(T_p)}{200} \right)^{ \scalebox{1.01}{$\frac{1}{6}$} }.
\end{equation}
In Fig.~\ref{fig:GW_DPDM}, we show the contours of the required value of $m_{A'}$. If $m_{A'}$ is instead taken smaller than that in Eq.~(\ref{eq:DPDM_mass}), e.g., to avoid warm dark matter, the dark photon contributes to dark radiation as in Eq.~(\ref{eq:DeltaNeff}).

\subsection{Effects of scattering}
\label{sec:DPDM_scatt}
Although the coupling between the axion and the dark photon in Eq.~(\ref{eq:axion-DP}) is suppressed by a loop factor, the dark gauge coupling, and a decay constant, dark photons and axions can still interact efficiently because of the enormous number densities involved. Using kinetic theory~\cite{Zakharov:1985,Zakharov:1992,Micha:2004bv}, which is essentially a perturbative computation that incorporates the relevant Bose enhancement, the scattering rate 
involving $N_i$ initial and $N_f$ final state particles with a typical momentum $k$ is
\begin{equation}
\label{eq:rate}
\Gamma_{N_i{\rm -to-}N_f}(k) \sim \left( \frac{\alpha_D}{4 \pi} \right)^{2N_V} \frac{f_{k}^{N_V} k^{2N_V+1}}{S^{2N_V}}.
\end{equation}
Here $N_V = N_i+N_f-2$ is the number of axion-dark photon vertices involved in the process. $f_k$ is the occupation number of particles at the momentum $k$. Just after the production, $k\sim k_{\rm TI}\sim \alpha_D \dot{\theta} / 4\pi$ and $f_k \sim (4\pi/\alpha_D)^4 S^2 / \dot{\theta}^2$.  Thus, the rate  $\Gamma_{N_i{\rm -to-}N_f} \sim k_{\rm TI} \sim H_p$, so the scattering is fact expected to be efficient for any $N_V$, and it is unclear which $N_{V}$ dominates.

A linear combination of the $U(1)_P$ charge in the axion rotation and the helical density of the dark photon is conserved. This conserved charge is analogous to the conserved particle number in a cold atom system, and it allows the scattering to evolve the system towards the formation of a Bose-Einstein Condensate (BEC) via particle cascade~\cite{Blaizot:2011xf,Berges:2012us}. It is conceivable that the BEC--and its associated charge--could be in the form of the axion rotation, in which case dark photon dark matter would not be viable.  However, for a fixed $U(1)_P$ charge, the free-energy density of the system is smaller if the charge is stored in small momentum modes of helical dark photons rather than the rotating axion. Thus, we expect that the charge is predominantly stored in the form of helical dark photons. As the system evolves towards a BEC, these dark photons cool, and the dark matter warmness constraint may be relaxed relative to the case without scattering presented in Sec.~\ref{sec:DPDM_no_scatt}. Since the helicity density is conserved, the prediction on the dark photon mass remains the same.

A numerical lattice computation is necessary to follow the evolution of the dark photons and axions. As the system evolves towards low momenta, the dark photon occupation number increases, and the rate in Eq.~(\ref{eq:rate}) becomes larger for larger $N_V$; a perturbative computation based on the kinetic theory fails. If the scattering rate is suppressed at low momenta by a non-perturbative effect, which happens in the case of real scalar fields with a quartic potential~\cite{Berges:2008wm,Berges:2012us}, evolution towards IR may be suppressed. Also, the correlation length of the would-be BEC should be limited by causality and stays under the horizon size, and hence the momentum of the dark photons cannot be arbitrarily small. Because of these factors, an observable degree of warmness of dark matter may remain. A lattice simulation can clarify this prediction.

We have argued above for a decrease in the typical momentum of the helical dark photons via scattering.  However, the total energy density of the axion/dark photon system should be conserved.  This necessitates the production of a component of the axions/dark photons that does not carry charge, which we refer to as non-helical.
The fate of these non-helical photons crucially depends on whether number-changing processes are efficient for these photons.  If they are efficient for all momenta, the non-helical component will evolve towards both kinetic and chemical equilibria, and  if achieved, have a typical comoving momentum much larger than $k_{\rm TI}$ and a number density much smaller than that of the helical component. Thus, if the evolution towards kinetic and chemical equilibria is sufficient, then the dark matter resides in the helical component, with the non-helical dark photons comprising a small contribution to dark radiation. On the other hand, if number-changing processes are ineffective (at least at low momenta), the non-helical component can also evolve towards a BEC.
This would be perhaps surprising, given the estimate of Eq.~(\ref{eq:rate}) that shows the importance of all higher order interactions.  However, given the importance of non-perturbative effects, as referenced above, we cannot rule out this possibility.  In fact, this behavior has been observed in a lattice simulation  of relativistic real scalar fields with a quartic potential~\cite{Berges:2012us}. To conclusively settle the fate of this non-helical component, a lattice simulation is necessary.

We emphasize that the number density of the non-helical component is not conserved as a whole. 
This opens the possibility that the number density of the non-helical component could in fact dominate the helical one.  For this to occur while satisfying energy conservation, number-increasing processes would need to be effective either at the initial stage (before the evolution towards a BEC occurs) or in the high momentum regime while the evolution towards a BEC occurs in a low momentum regime.  If the dark matter is primarily provided by these non-helical dark photons, the prediction for its mass becomes smaller. The mass in Eq.~(\ref{eq:DPDM_mass}), which was predicted by the helical density, can then be understood as an upper bound on the dark photon mass. Because of the cooling of the non-helical dark photons during the evolution towards the BEC by number-conserving processes, the warmness constraint can be relaxed compared to that from Eq.~(\ref{eq:DPDM_warmness}).

If instead the dark photon is massless, it contributes to dark radiation. Despite the possible scattering, the estimation in Eq.~(\ref{eq:DeltaNeff}) is generic. It still applies as long as the axion does not continue to produce dark photons after $T_p$.

\section{Concrete Realization} 
\label{sec:model}

We have thus far discussed the production of dark photons and gravitational waves in as model-independent a fashion as possible. However, it is of interest to construct a concrete model and discuss a complete cosmology as an existence proof.  

We explore the parameter space motivated by the potential GW signal reported by NANOGrav~\cite{Arzoumanian:2020vkk} and also for a smaller signal strength within the reach of SKA~\cite{Janssen:2014dka}.  We then discuss a possible future GW signal at much higher frequencies, e.g., relevant for LISA and DECIGO/BBO, and discuss implications for dark matter in that part of the parameter space.   In all cases, there are two particularly important questions: the approximate form of the saxion potential and the mechanism by which the saxion is thermalized.

\subsection{Scalar potential}
\label{sec:potential}

To produce an observable amount of GWs, the energy density contained in the axion rotation  must be large at the time of tachyonic instability. 
The rotation of the complex field $P$ begins when the Hubble expansion rate is comparable to the saxion mass. 
Since the energy density does not redshift until the onset of the rotation, a smaller saxion mass, namely, a flatter saxion potential, is advantageous for enhancing GW signals.

We first discuss the possibility that the saxion potential is approximately quartic at large field values, $\lambda^2 |P|^4$, as might be expected in the case where the theory does not possess supersymmetry. We will show that a successful tachyonic instability is difficult in this case.

For an initial field value $S_i$, the initial saxion mass in the quartic potential, $\sim \lambda S_i$, is bounded by the Hubble scale during inflation $H_{I}$. The bound on the tensor to scalar ratio from the CMB then imposes $\lambda S_i< 6\times10^{13}$ GeV ~\cite{Akrami:2018odb}. For a given strength of a GW signal, this translates to a bound on the quartic coupling as $\lambda \lesssim 10^{-3} (10^{-12}/\Omega_{\rm GW} h^2)^{1/4} (20/r_p)^{1/2} $. However, for a given value of the quartic coupling $\lambda$, the coupling $y_{\psi}$ between the saxion and $U(1)_P$-charged fermions  $y_{\psi} P \bar{\psi} \psi$ is bounded by the requirement that there be no excessive quantum correction to $\lambda$.  Therefore, we expect $y_{\psi}^4 \lesssim 16 \pi^2  \lambda^2$.  This bound on $y_{\psi}$  in turn  imposes an upper bound on the mass of the $U(1)_P$-charged fermions at the time of the tachyonic instability. This bound is sufficiently strong that $U(1)_P$-charged fermions are unavoidably kinematically accessible at the time of production, $m_{\psi} <T_{p}$, and there is a danger that there is a thermal population of these fields.   

A thermal population of such fermions is problematic because their presence leads to a non-trivial dark electric conductivity of the universe~\cite{Baym:1997gq,Arnold:2000dr,Domcke:2018eki}, which acts as a damping force in the dark photon equation of motion.  This impedes the amplification needed for an effective tachyonic instability.  One might have thought that tiny $y_{\psi}$ might allow only a sufficiently small abundance of the fermions to be produced, and hence a sufficiently small conductivity.  However, $y_{\psi}$ cannot be taken vanishingly small. Most importantly, a sufficiently large fermion mass is required to suppress the production of fermions via the analog of the Schwinger effect~\cite{Domcke:2021fee}.   We find the lower bound of on $y_{\psi}$ coming from this consideration leads to an excessive freeze-in abundance of the fermions, and hence an excessive electric conductivity. This consideration is robust, even in the potential presence of a large self-interaction amongst the $U(1)_P$-charged fermions that reduces the conductivity.

For this reason, we now concentrate on a supersymmetric scenario, where the flatness of the saxion potential is radiatively stable. We focus on the case where the saxion has an approximately quadratic potential with a typical mass $m_S$. 
This can be realized in a model with a global $U(1)_P$ symmetry spontaneously broken by dimensional transmutation from the running of the soft mass~\cite{Moxhay:1984am}, 
\begin{equation}
\label{eq:logV}
V = m_S^2 |P|^2 \left( {\rm ln}\frac{2 |P|^2}{f_\phi^2} -1  \right),
\end{equation}
and can also be realized in a supersymmetric two-field model with soft masses,
\begin{equation}
\label{eq:two_field}
W = \lambda X \left( P \bar{P}- V_P^2 \right), ~~V_{\rm soft} = m_P^2 |P|^2 + m_{\bar{P}}^2 |\bar{P}|^2,
\end{equation}
where $X$ is a chiral multiplet whose $F$-term fixes the global symmetry breaking fields $P$ and $\bar{P}$ along the moduli space $P \bar{P} = V_P^2$.

The fermionic superpartner of the axion, the axino, obtains a mass from supersymmetry and/or $R$ symmetry breaking. In the model described by Eq.~(\ref{eq:two_field}), the tadpole term for $X$ generated via supergravity effects, $\lambda X V_P^2 m_{3/2}$, induces a vacuum expectation value $\vev{X}\sim m_{3/2}/\lambda $, and the axino obtains a Dirac mass $\sim m_{3/2}$. When the saxion is thermalized, axinos are also thermalized; this leads to production of axinos. 
For the relevant range of gravitino masses discussed later, the thermalized axinos would have to be diluted by entropy production, e.g., from the decay of long-lived particles, to avoid overclosure.
In this case, the analysis on the thermalization of the rotation in Sec.~\ref{sec:parameter_space} should be modified by taking into account the dilution. 

For sufficiently large $m_{3/2}$, we may instead introduce $R$-Parity violation so that the axino with a mass $\sim m_{3/2}$ can decay into Standard Model particles. And for $m_{3/2} > {\cal O}(1)$ TeV, there is yet another possibility: the axino may decay into the Lightest Observable-sector Supersymmetric Particle (LOSP).  Notably, this case is compatible with gravity mediation for the visible sector soft masses. As we will see in Sec.~\ref{sec:parameter_space}, such a large $m_{3/2}$ is consistent with a high-frequency region motivated by LISA, DECIGO, and BBO, but not with a low-frequency region motivated by NANOGrav and SKA. We comment on these possibilities again in Sec.~\ref{sec:baryogenesis}, where we discuss baryogenesis from the axion rotation.

However, for simplicity, we will focus on the model in Eq.~(\ref{eq:logV}), where the axino may remain sufficiently light that even its thermal abundance may not cause any conflict with observations, and no dilution is required.   In this model, the axino mass is generated by a one-loop quantum correction from the coupling of $P$ with $\bar{\psi}\psi$ in Eq.~(\ref{eq:mediator}) and is given by
\begin{align}
\label{eq:axino_ypsi}
    m_{\widetilde{a}}\simeq \frac{y_\psi^2 A_\psi}{16\pi^2} ,  \qquad \qquad \qquad \rm{(dimensional \; transmutation \; potential)}
\end{align}
where $A_\psi$ is the scalar trilinear term associated with the coupling, $V \supset y_\psi A_\psi P \widetilde{\bar{\psi}}\widetilde{\psi}$. In this expression, we have assumed the mass of $\bar{\psi}\psi$ is dominated by the supersymmetric mass term $y_\psi P$. At minimum, the $A$ term receives a contribution from anomaly mediation~\cite{Randall:1998uk,Giudice:1998xp}, $A_\psi\simeq y_{\psi}^2m_{3/2}/(16\pi^2)$.%
\footnote{A coupling between the supersymmetry breaking field $Z$ and $P$, $\psi$, and $\bar{\psi}$ in the K\"{a}hler potential, e.g., $PP^\dag ZZ^\dag$, also generates an $A$ term after $Z$ obtains a vacuum expectation value, as required in gauge mediation, but we find this contribution is typically negligible.}
The resultant axino mass is given by
\begin{equation}
    \label{eq:axinomassnumerical}
    m_{\widetilde{a}}\simeq \frac{y_{\psi}^4 m_{3/2}}{(4\pi)^4} \simeq 4~{\rm eV} \times  y_{\psi}^4 \left(\frac{m_{3/2}}{100 \keV}\right).
\end{equation}
Assuming that the axino decouples from the thermal bath when the number of effective degrees of freedom of the bath is around $100$, the upper bound on the axino mass is $m_{\widetilde{a}} < 4.7$ eV~\cite{Osato:2016ixc}.
To achieve this light axino, $y_\psi$ and/or $m_{3/2}$ must be sufficiently small.

However, the coupling $y_\psi$ cannot be arbitrarily small. The potential in Eq.~(\ref{eq:logV}) in terms of the parameters of theory is
\begin{equation}
    V = m_P^2 |P|^2 + \frac{y_\psi^2 m_{\widetilde{\psi}}^2 } {8 \pi^2} |P|^2 \log \left(\frac{y_\psi^2|P|^2}{\mu^2} \right),
\end{equation}
where $m_P$ and $m_{\widetilde{\psi}}$ are the soft masses of $P$ and $\psi/\bar{\psi}$ at a UV scale $\mu$. 
The second term is the quantum correction that arises from loops of the $U(1)_P$ field $\psi$. 
To successfully break the $U(1)_P$ global symmetry by dimensional transmutation, the soft mass must be smaller than the supersymmetric mass, i.e., $m_{\widetilde{\psi}} < y_\psi f_\phi$. Furthermore, the saxion mass is dominated by the quantum correction so $m_S \simeq y_\psi m_{\widetilde{\psi}} / 4 \pi$. Taken together, these conditions imply 
\begin{equation}
    y_\psi^2 f_\phi > 4\pi m_S.
    \label{eq:masslimit}
\end{equation}
For simplicity, we approximate the saxion potential at $S > f_\phi$ by an exactly quadratic one with a mass $m_S$, ignoring the logarithmic correction (except for the discussion on the Q-ball formation by a thermal potential around Eq.~(\ref{eq:Q_ball})), but we will impose the condition of Eq.~(\ref{eq:masslimit}). 

Successful dimensional transmutation also imposes a constraint on the underlying soft supersymmetry breaking parameters of the ultraviolet theory.  Since $y_\psi$ is bounded from above to suppress the quantum correction to the axino mass in Eq.~(\ref{eq:axinomassnumerical}), and $m_P^2 \sim - y_\psi^2 m_{\widetilde{\psi}}^2 {\rm log}(y_\psi^2 f_\phi^2/\mu^2)/(8\pi^2)$, the soft mass $m_{\widetilde{\psi}}^2$ must be much larger than $m_P^2$ for a large gravitino mass. For example, for $m_{3/2} \sim 1$ GeV, $m_{\widetilde{\psi}} > 10 m_{P}$ is required.  This could be achieved by a modest hierarchy in the suppression scales of the couplings of the supersymmetry breaking field with $\psi$ and $P$.

To give the axion a rotational kick, we consider an explicit $U(1)_P$ symmetry breaking superpotential term,
\begin{align}
    W = \frac{1}{n}\frac{P^n}{\Lambda^{n-3}},
\end{align}
where $\Lambda$ is a cut off scale. The scalar potential of $P$ is then given by
\begin{align}
\label{eq:saxion potential}
    V = m_S^2|P|^2 + c_H H^2 |P|^2+ \frac{|P|^{2n-2}}{\Lambda^{2n-6}} + \left(A \frac{P^n}{\Lambda^{2n-6}} + {\rm h.c.}\right),
\end{align}
where $H$ is the Hubble scale, $c_H$ is a dimensionless parameter, and $A$ is a dimensionful parameter. In the case that supersymmetry breaking is mediated to $P$ via supergravity, we expect scalar masses and $A$-terms to both be of order the gravitino mass $m_{3/2}$.

We assume that $c_H$ during inflation is negative and ${\mathcal O}(1)$, or $|c_H| \ll 1$.
For the former case, the field value of the saxion $S \equiv \sqrt{2}|P|$ during inflation is fixed at a large value with $V'(S)=0$~\cite{Dine:1995uk,Dine:1995kz}, where the second and the third terms of Eq.~(\ref{eq:saxion potential}) balance with each other.  For the latter case, assuming that the initial field value of the saxion is large, e.g., around the fundamental scale, the saxion field value follows a classical attractor solution with $V''(S)\sim H^2$~\cite{Harigaya:2012up}.%
\footnote{If inflation lasts for a long time, the saxion field value follows a distribution determined by quantum fluctuations~\cite{Vilenkin:1982de, Vilenkin:1982wt, Linde:1982uu, Starobinsky:1982ee}.}
The angular field value may be initially random but is homogenized by inflation.
We assume that $c_H$ \emph{after} inflation also has the property described above. 
After inflation, as the Hubble scale decreases, the saxion field value tracks a point with $V''(S) \sim H^2$~\cite{Dine:1995uk,Dine:1995kz,Harigaya:2015hha}. Once the Hubble scale becomes smaller than $m_S$, the $P$ field begins rotations with a kick induced by the $A$-term. One can show that the ratio between the potential gradient in the angular direction and that in the radial direction is $\mathcal{O}(A/m_S)$ (e.g., Ref.~\cite{Co:2020jtv}). If $A\sim m_S$, as expected  if the supersymmetry breaking soft terms of $P$ are mediated by gravity, the rotations have ${\mathcal O}(1)$ ellipticity. In what follows, we assume $m_{3/2} \lesssim m_S$ to achieve ${\mathcal O}(1)$ ellipticity. Since the field value of $P$ is homogenized by inflation, the rotation is nearly coherent.

\subsection{Allowed parameter space}

\begin{figure}
    \includegraphics[width=0.495\columnwidth]{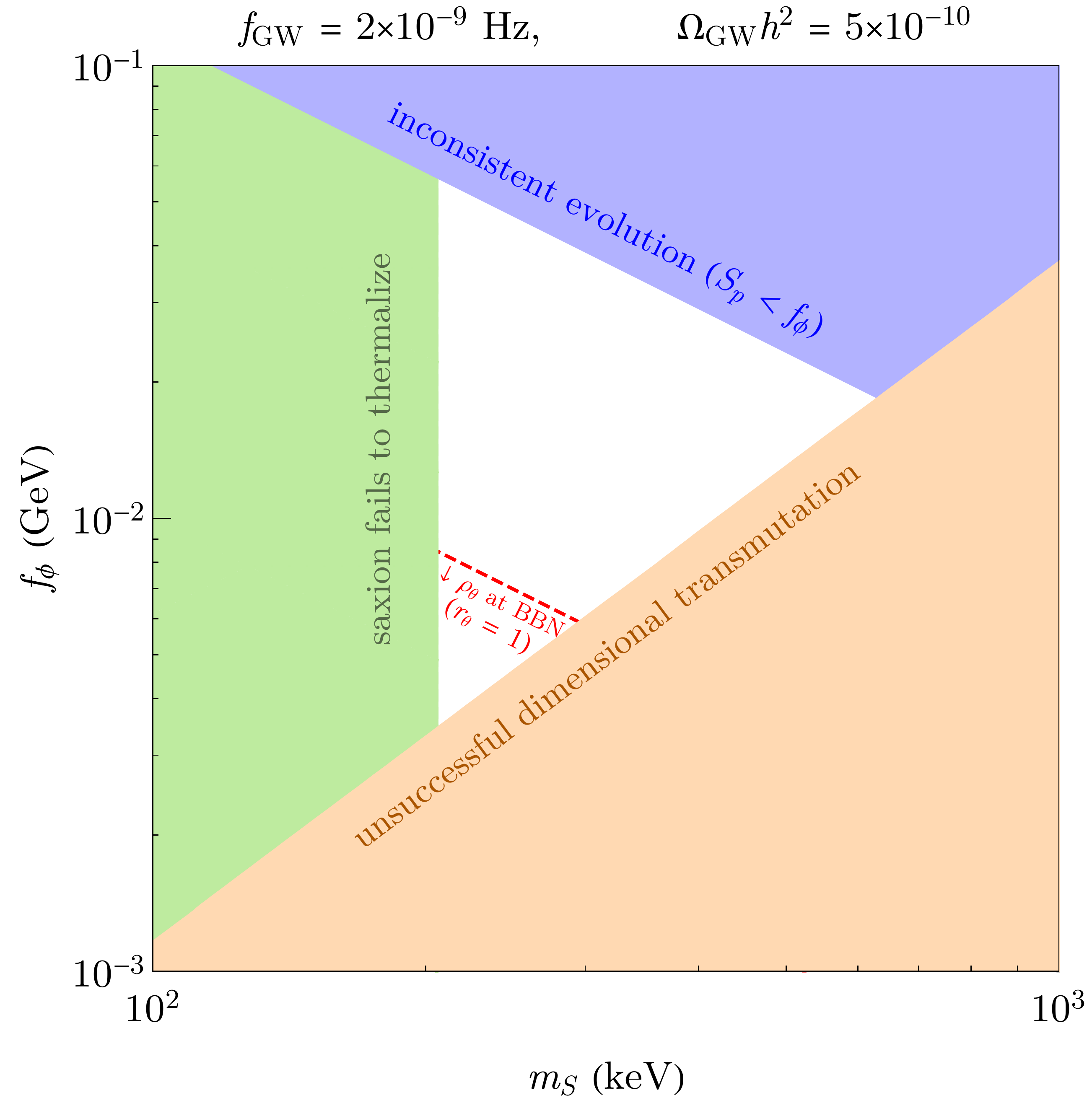}
    \includegraphics[width=0.495\columnwidth]{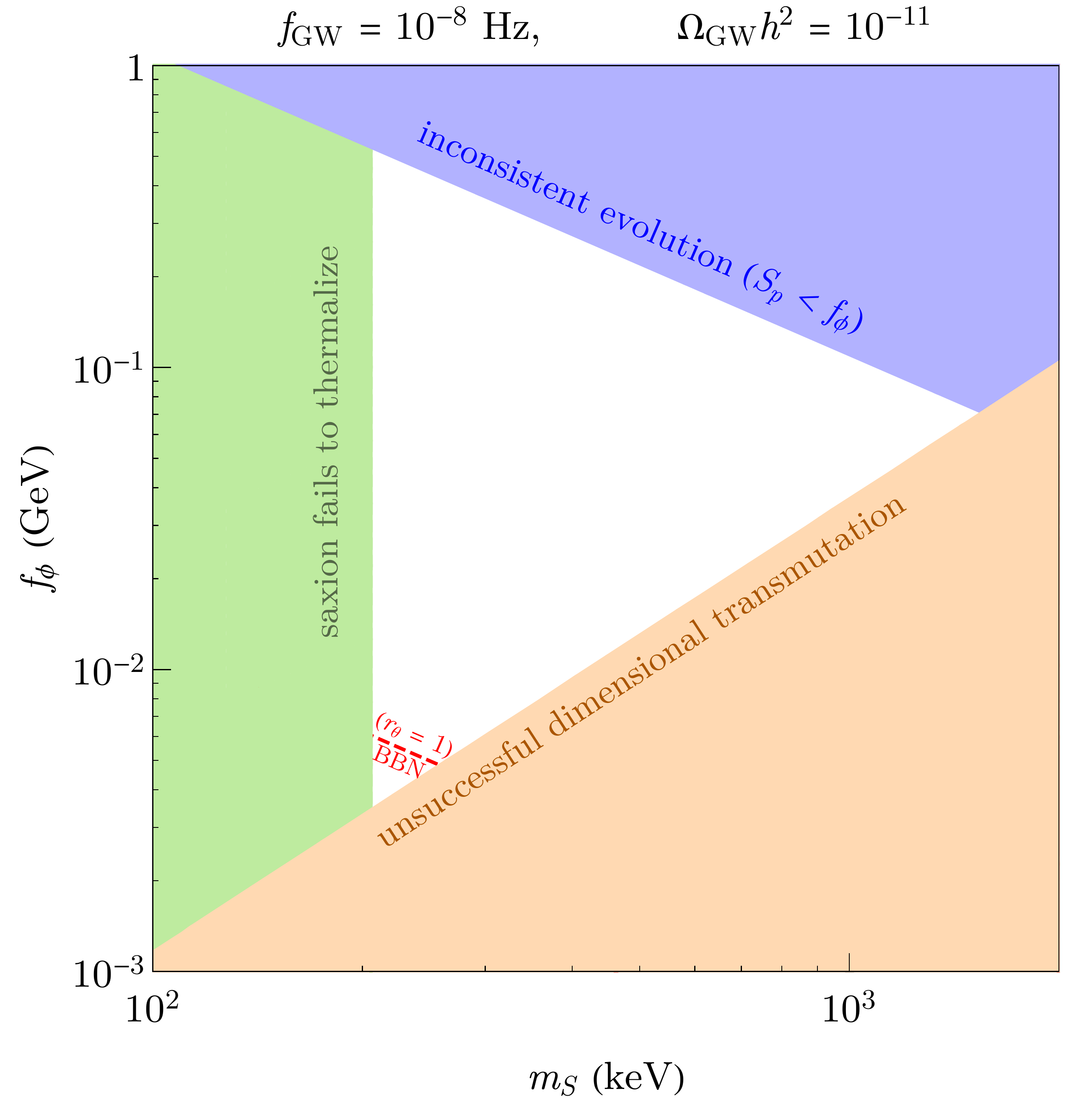}
    \includegraphics[width=0.495\columnwidth]{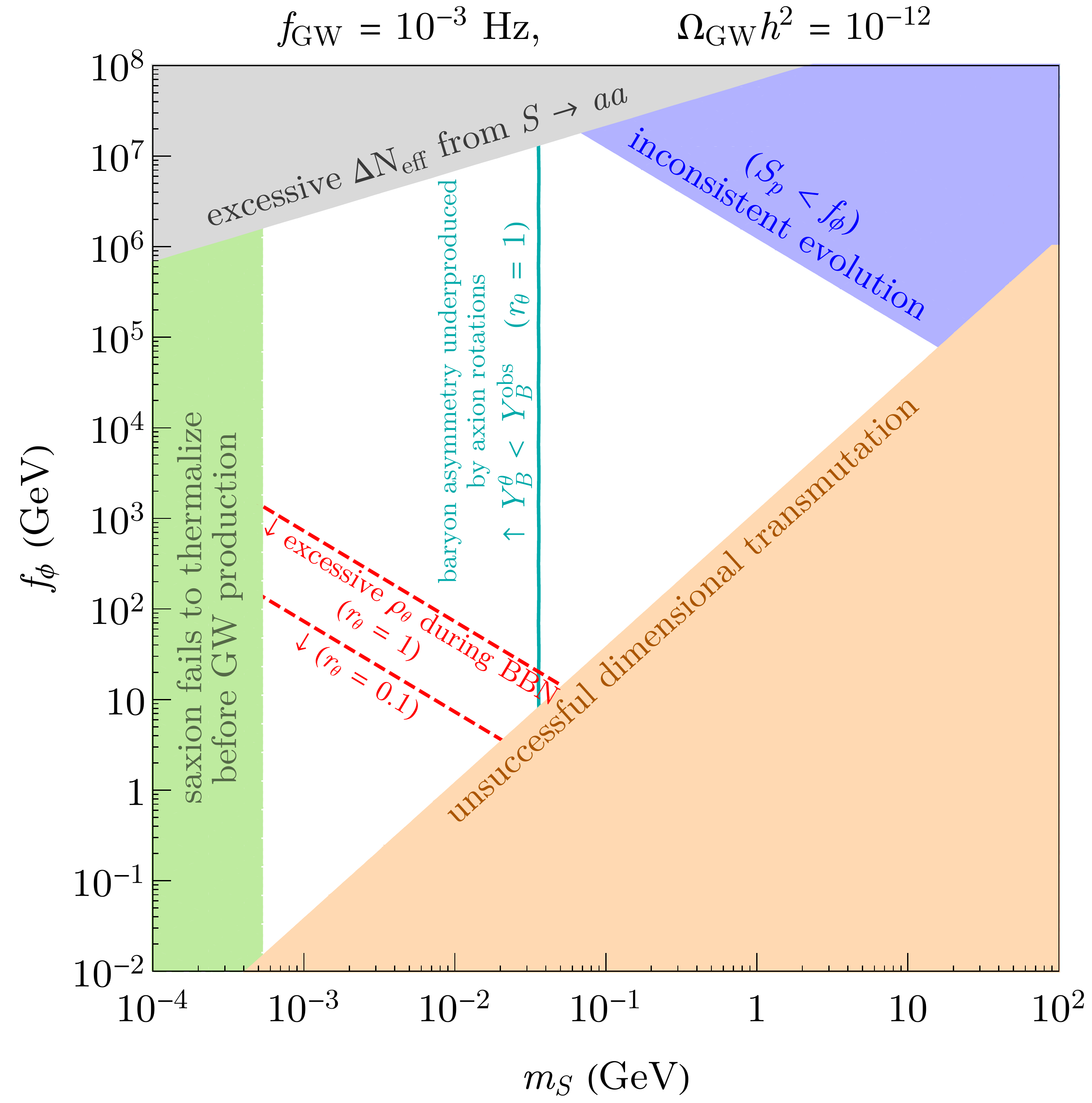}
    \includegraphics[width=0.495\columnwidth]{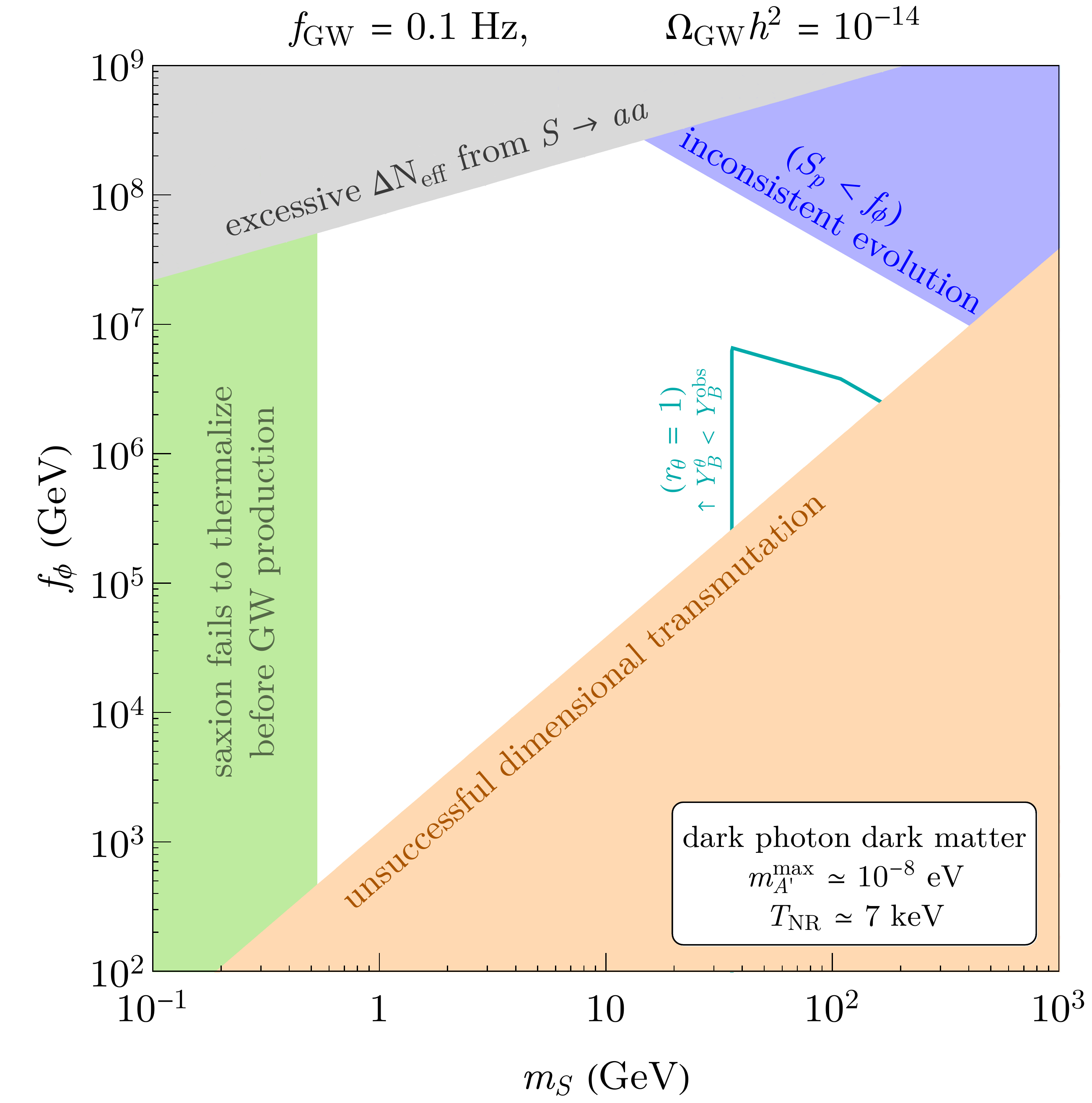}
    \caption{The white regions may lead to GW signals as motivated by NANOGrav (top left), SKA (top right), LISA (bottom left), and DECIGO and BBO (bottom right). We fix $m_L = 1 \TeV$ and $r_p = 20$ with $g_*(T_p) = 10.75$ (200) for the top (bottom) panels. The exclusion regions follow from Eq.~(\ref{eq:masslimit}) for orange, below Eq.~(\ref{eq:rhorot}) for blue, above (around) Eq.~(\ref{eq:s_aa}) for red (gray), and Eqs.~(\ref{eq:yLmin_therm}) and (\ref{eq:Q_ball}) for green. To the right/below the cyan lines, the axion rotation can also explain the observed baryon asymmetry via Eq.~(\ref{eq:yL_YB}). In the bottom right panel, the dark photons may explain dark matter. In other panels, they can be sufficiently cold to be dark matter if a cascade to IR occurs via scattering, as considered in Sec.~\ref{sec:DPDM_scatt}; otherwise, they are taken light, contributing to $\Delta N_{\rm eff}$ by Eq.~(\ref{eq:DeltaNeff}).}
    \label{fig:model}
\end{figure}

The main result of the paper is shown in Fig.~\ref{fig:model}.  In the white regions, the axion rotation, combined with the tachyonic instability, can generate GW signals with the frequency and magnitude as indicated in each panel. The benchmark point of $f_{\rm GW}$ and $\Omega_{\rm GW}h^2$ in the top left panel of Fig.~\ref{fig:model} is motivated by the potential NANOGrav signal~\cite{Arzoumanian:2020vkk}, while those for other three panels can be reached by SKA~\cite{Janssen:2014dka} (top right), LISA~\cite{Audley:2017drz} (bottom left), and DECIGO~\cite{Sato:2017dkf} and BBO~\cite{Crowder:2005nr} (bottom right). The constraints indicated by the blue, orange, and gray shading as well as red dashed lines are independent of thermalization models and are explained in Sec.~\ref{sec:various}. The green shading is the constraint for a thermalization model with a vector-like lepton with a mass $m_L$ and is discussed in Sec.~\ref{sec:parameter_space}. The cyan lines show the parameter region that is compatible with baryogenesis from the axion rotation and is discussed in Sec.~\ref{sec:baryogenesis}.

In all panels of Fig.~\ref{fig:model} except the bottom right, dark photon dark matter is too warm (independent of the thermalization model) as indicated by the solid red curve in Fig.~\ref{fig:GW_DPDM}, unless dark photons cascade into IR via  dark photons/axions scattering from the interaction in Eq.~(\ref{eq:axion-DP}), as discussed in Sec.~\ref{sec:DPDM_scatt}.
In these cases, we may assume the dark photon mass is significantly smaller than that in Eq.~(\ref{eq:DPDM_mass}) so that the dark photon contributes to dark radiation at a level given by Eq.~(\ref{eq:DeltaNeff}). Then another dark matter candidate is needed; one example is the QCD axion.%
\footnote{One might have thought that the $U(1)_{P}$-charged fermion $\psi$ could be the dark matter, but we find that the combination of achieving a successful freeze-out abundance and efficient tachyonic instability for the dark photon are incompatible in almost all of the viable parameter space, see Appendix~\ref{app:concrete_cosmo} for details.}
Furthermore, due to small $f_\phi$ in the top panels, the $U(1)_P$-charged fermions $\psi$ are necessarily light, and they may have millicharge, $Q_\psi$, via a kinetic mixing $\chi$.   The supernova constraint~\cite{Chang:2018rso} requires $Q_\psi \lesssim 10^{-9}$ for $m_\psi \lesssim 10 \MeV$. However, given that the millicharge $Q_\psi \simeq \chi e_D/e_{\rm EM}$ is already suppressed due to small $e_D$ in Eq.~(\ref{eq:gaugecoupling}), the resultant bound $\chi \lesssim 0.05 ({\rm nHz} / f_{\rm GW})(m_S / 300 \keV)^{1/2} $ is very weak. 

On the other hand, in the bottom right panel, dark photon dark matter is consistent with the warmness constraint shown in Fig.~\ref{fig:GW_DPDM} for a GW strength within the reach of DECIGO~\cite{Sato:2017dkf} and BBO~\cite{Crowder:2005nr}. Future  21-cm surveys~\cite{Sitwell:2013fpa} can probe the entire parameter space in this panel via the warmness of dark matter, though the robustness of this conclusion depends on a more detailed understanding of the scattering effects of Sec.~\ref{sec:DPDM_scatt}. A dark photon in this mass regime could be probed by the DM Radio experiment~\cite{Chaudhuri:2014dla}, assuming that the kinetic mixing $\chi$ with the Standard Model were sufficiently large, $\chi \gtrsim 10^{-16}$. For $m_{A'} \simeq 10^{-8} \eV$ as a maximum value of $m_{A'}$ predicted in Eq.~(\ref{eq:DPDM_mass}), $\chi \lesssim 10^{-12}$ is needed to avoid distortions of the CMB spectrum~\cite{Arias:2012az}. The supernova constraint~\cite{Chang:2018rso} on $\chi$ via the millicharged $\psi$ can be avoided since $m_\psi$ can be made larger than $\mathcal{O}(100) \MeV$.

The remainder of this section is devoted to explaining the origin of the exclusion regions.

\subsubsection{Constraints independent of thermalization model}
\label{sec:various}

From Eqs.~(\ref{eq:fGW_today}) and (\ref{eq:GWSignal}), the energy density of the rotating field may be related to the GW signal via
\begin{align}
\label{eq:rhorot}
\rho_\theta (T_p) 
& \simeq (3 \MeV)^4 \left( \frac{f_{\rm GW}}{2~{\rm nHz}} \right)^4 \left( \frac{\Omega_{\rm GW}h^2}{5 \times 10^{-10}} \right)^{ \scalebox{1.01}{$\frac{1}{2}$} } \left( \frac{g_* (T_p)}{10} \right)^{ \scalebox{1.01}{$\frac{1}{2}$} } \left( \frac{20}{r_p} \right)^3 \\
& \simeq (60 \TeV)^4 \left( \frac{f_{\rm GW}}{0.1~{\rm Hz}} \right)^4 \left( \frac{\Omega_{\rm GW}h^2}{10^{-14}} \right)^{ \scalebox{1.01}{$\frac{1}{2}$} } \left( \frac{g_* (T_p)}{200} \right)^{ \scalebox{1.01}{$\frac{1}{2}$} } \left( \frac{20}{r_p} \right)^3 , \nonumber
\end{align}
where the first (second) line is normalized to values relevant for NANOGrav (DECIGO/BBO).
The temperature at production is computed from Eq.~(\ref{eq:Tprod}) and our benchmarks are given by $T_{p} \simeq 5.5$ MeV for NANOGrav, $T_{p} \simeq 27$ MeV for SKA, $T_{p} \simeq 1.7$ TeV for LISA, and $ T_{p} \simeq 170$ TeV for DECIGO/BBO.
Since the saxion field value at production is always greater than or equal to its value at the minimum of the potential, $S_p \ge f_\phi$, and $\dot\theta \simeq m_S$, the expression Eq.~(\ref{eq:rhorot}) taken  with $\rho_\theta(T_p) = \dot{\theta}_p^2 S_p^2$  bounds $f_\phi$ as a function of $m_S$.  This is shown by the blue regions of Fig.~\ref{fig:model}. 

While $m_S \le f_\phi$ is required for the perturbativity of the saxion potential, a stronger condition is provided by Eq.~(\ref{eq:masslimit}).  The precise relation between   $f_{\phi}$ and $m_S$ depends on $y_{\psi}$.  This in turn will depend upon the requirement of avoiding fine-tuning in the axino mass, as in Eq.~(\ref{eq:axinomassnumerical}).  This requirement excludes the orange regions of Fig.~\ref{fig:model}. Taken together, these constraints show that the saxion must be light.

After the production of dark photons by tachyonic instability, the axion field may continue rotating. Its energy density behaves as matter for $S>f_\phi$ and kination for $S\simeq f_\phi$ according to Eqs.~(\ref{eq:scaling_matter}) and (\ref{eq:scaling_kination}), respectively.  The energy density of the axion rotation should be subdominant throughout BBN to prevent modification of the expansion rate of the universe. Below the red dashed lines in Fig.~\ref{fig:model}, the energy density of the axion rotation is larger than the equivalent of $\Delta N_{\rm eff} = 0.4$  for some temperature between $T=1$ MeV and $0.1$ MeV, and we expect light elements abundances may be modified to an unacceptable level.   Here we show this constraint for $r_\theta = 1$ and $0.1$ in Fig.~\ref{fig:model}, where $r_\theta \le  1$ is defined in Eq.~(\ref{eq:rtheta}) and parameterizes the residual amount of $U(1)_P$ charges in axion rotations after the dark photon production by tachyonic instability.
We note that the upper bound on $N_{\rm eff}$ during BBN dominantly comes from its effect on the neutron-proton conversion around $T\sim$ MeV and its impact on the helium abundance. When the axion rotation behaves as matter between $T=1$ MeV and $0.1$ MeV our requirement that the axion energy density be subdominant during this entire period comes from the lowest temperature,   
$T=0.1$ MeV, by which time the neutron-proton conversion is largely complete. In this case, our constraint is likely somewhat too strong; a more precise determination of the constraint is beyond the scope of this paper.
In deriving such a bound, the contribution to the energy density from thermalized axions, saxions, and axinos ($\Delta N_{\rm eff} \simeq 0.1$) and dark photons produced by tachyonic instability in Eq.~(\ref{eq:DeltaNeff}) should be also included.

The thermalized saxions will eventually decouple and then decay to axions with a rate
\begin{equation}
\label{eq:s_aa}
    \Gamma_{S \rightarrow aa} = \frac{m_S^3}{32 \pi f_\phi^2}.
\end{equation}
The contribution to dark radiation is negligible if the decay occurs before the decoupled saxions become non-relativistic. The gray region in the bottom panels of Fig.~\ref{fig:model} is excluded because the decay happens only after the saxions become non-relativistic and the dark radiation abundance is enhanced.

The tachyonic instability can be obstructed if there exists a dark electric conductivity for the dark photons provided by dark-charged particles. While the precise dark content is model-dependent, at minimum, there exist particles $\psi$ that appear in the Lagrangian with coupling $y_\psi P \bar{\psi} \psi$ and are integrated out at low energies to realize the interaction in Eq.~(\ref{eq:axion-DP}).  If these particles are sufficiently heavy as not to be thermally populated at $T_p$, then this dangerous dark electric conductivity for the dark photon is avoided. In the entire parameter space of interest, $\psi$ particle's mass, $m_\psi = y_\psi S(T)$, can be made sufficiently large by using a Yukawa coupling $y_\psi$ less than unity and consistent with the upper bound from the quantum correction to the axino mass in Eq.~(\ref{eq:axino_ypsi}).

\subsubsection{Thermalization constraints}
\label{sec:parameter_space}

We consider saxion thermalization via the scattering of $P$ with new vector-like leptons. 
We add a pair of vector-like Standard Model charged leptons $L$ and $\bar{L}$ as follows:
\begin{equation}
\label{eq:super_vector}
    W = (y_L P \ell + m_L L) \bar{L},
\end{equation}
where $y_L$ and $m_L$ are a Yukawa coupling and a vector-like mass, respectively, and $\ell$ is a Standard Model lepton doublet. It is also possible to introduce the vector-like fermions in a complete multiplet of a grand unified theory to maintain unification.
The global $U(1)_P$ charges are $L (-1)$ and $\bar{L} (+1)$ and the $U(1)_P$ symmetry does not have a mixed anomaly with the Standard Model gauge symmetry, which loosens observational constraints on $f_\phi$. The coupling $y_L$ will be responsible for the thermalization of the saxion field. When the vector-like fermions are in the thermal bath, the thermalization rate of $P$ is
\begin{equation}
\label{eq:therm_rate}
    \Gamma_{\rm th} \simeq \alpha_2 y_L^2 T,
\end{equation}
which decreases more slowly than the Hubble expansion rate does. We require thermalization to occur before $L$ falls out of thermal equilibrium, i.e.,~$T=m_L$, 
and also before the production of dark photons by tachyonic instability occurs, $T=T_p$.  This ensures that the motion of the complex field $P$ becomes nearly circular and coherent at the time of instability; the possibility of relaxing this requirement is discussed in Sec.~\ref{sec:thermalizationComm}. This places a lower bound on $y_L$,
\begin{equation}
\label{eq:yLmin_therm}
    y_L \gtrsim 2 \times 10^{-7} 
    \left( \frac{T_{\rm th}}{\rm TeV} \right)^{\scalebox{1.01}{$\frac{1}{2}$} } 
    \left( \frac{g_*(T_{\rm th})}{100} \right)^{\scalebox{1.01}{$\frac{1}{4}$} } ,
\end{equation}
with the thermalization temperature $T_{\rm th} = \max(m_L, T_p)$.%
\footnote{Another lower bound on $y_L$ could arise from avoiding parametric resonance production of the scalar partner of $L$ during the non-circular rotation of $P$. This can be avoided if $y_L S > m_S$, so that the mass of the scalar partner oscillates adiabatically. However, the scalar partner obtains a thermal mass $\sim T$, implying that $y_L S <T$ is sufficient to achieve adiabaticity. Since $T_{\rm th} > m_S$ in the relevant parameter region, whenever $y_L S > m_S$ is violated, $y_L S <T$ is satisfied.}
Production temperatures for the different benchmarks are given below Eq.~(\ref{eq:rhorot}).
The lower bound on $m_L$ from collider search at the LHC is 790 GeV when the dominant decay mode is to the third generation leptons~\cite{Sirunyan:2019ofn}.  For concreteness, we fix $m_{L} = 1$ TeV in our analysis.

While this constraint favors large $y_{L}$, there are a number of upper bounds of $y_{L}$ that we will discuss below.  Tension between these upper bounds and the lower bound from the thermalization requirement will lead to the green regions in Fig.~\ref{fig:model}.

The fermions responsible for thermalization must be thermally populated themselves. This means that they must have a mass lower than the temperature at thermalization, 
\begin{equation}
\label{eq:thermalbath}
y_L S(T_{\rm th}) < T_{\rm th}.
\end{equation}
If this condition is violated, the mass of $L$ is always larger than the temperature for $T > T_{\rm th}$ since $S\propto R^{-3/2}$ and $T\propto R^{-1}$.

When the vector-like fermions are in the thermal bath, $S$ obtains a thermal mass $\sim y_L T$. If the thermal mass is larger than the vacuum mass $m_S$, $P$ rotates in the presence of a thermal potential with $\dot{\theta} \sim y_L T > m_S$. The thermal potential is not exactly quadratic, but is flatter than a quadratic one. The consequence of such a potential is that Q-balls are formed~\cite{Coleman:1985ki,Kusenko:1997zq,Kusenko:1997si,Kasuya:1999wu,Dine:2003ax}. The Q-balls evaporate into an inhomogeneous configuration of $P$ once the thermal mass becomes subdominant. For evaporation to occur, the logarithmic correction to the nearly quadratic zero-temperature potential in Eq.~(\ref{eq:logV}), which makes the potential steeper than a quadratic one, is essential; otherwise, even a subdominant thermal potential makes the potential slightly shallower than a purely quadratic potential.  The evaporation occurs when either $y_L T < m_S$ or $T < m_L$. For the latter case, however, the inhomogeneous configuration is not subsequently thermalized since $L$ decouples from the thermal bath. For successful thermalization of these would-be inhomogeneities, we require that the saxion thermal mass become smaller than its vacuum mass before $L$ decouples. We further require that the Q-balls evaporate before dark photon production occurs so that tachyonic instability is not disturbed.%
\footnote{We expect that it may be possible to relax this condition. Since the thermal potential is of the form $T^4 F(S/T)$, where $F$ is some function, the radius of the Q-balls increases in proportion to the scale factor $R$ after their production, and the Q-balls are not localized. In this case, we expect that the average angular velocity $\vev{\dot{\theta}}$ is  given by $V'/S \sim y_L T$ and the tachyonic instability could still occur with this modified angular velocity.}
In summary, for the Q-ball constraint, we require that the thermal mass be smaller than $m_S$ when $T=\max(m_L,T_p)=T_{\rm th}$,
\begin{align}
\label{eq:Q_ball}
    y_L T_{\rm th} < m_S.
\end{align}

The coupling $y_L$ introduces quantum corrections to the axino and saxion masses.
The axino receives a mass contribution from a loop of the vector-like fermions,
\begin{equation}
\label{eq:AxinoMass_yL}
    m_{\widetilde{a}} \simeq \frac{y_L^4 f_\phi^2 A_L}{16\pi^2 m_L^2},
\end{equation}
where we assume that the masses of vector-like fermions and sfermions are dominated by the Dirac mass term $m_L$.
At minimum, we expect a contribution to the $A$ term from a loop containing the Standard Model gaugino mass, $A_L \sim g_{\rm SM}^2 m_{\lambda}/(16 \pi^2)$, and therefore 
\begin{equation}
\label{eq:AxinoMass_gaugino}
    m_{\widetilde{a}} \gtrsim 20 \eV \left( \frac{y_L}{10^{-3}} \right)^4 \left( \frac{f_\phi}{10^6 \GeV} \right)^2 \left( \frac{m_{\lambda}}{\rm TeV} \right) \left( \frac{1 \TeV}{m_L} \right)^2 .
\end{equation}
We take visible sector soft masses to be ${\mathcal O}(1)\TeV$.  
The upper bound on the axino mass from warm dark matter~\cite{Osato:2016ixc} (see the discussion below Eq.~(\ref{eq:axinomassnumerical})), $m_{\widetilde{a}} < 4.7$ eV, then translates to an upper bound on $y_L$. The saxion mass also receives a quantum correction 
\begin{equation}
\label{eq:Delta_mS2}
   \Delta m_S^2 \simeq \frac{y_L^2 \tilde m_{\widetilde{L}}^2}{16 \pi^2},
\end{equation}
where $\tilde m_{\widetilde{L}}$ is the soft mass of $\widetilde{L}$, and the requirement that the saxion mass not be fine-tuned,  $\Delta m_S^2 < m_S^2$, also leads to an upper bound on $y_L$.

Two-loop corrections from the vector-like fermions and weak or hypercharge gauge bosons necessarily introduce the following flavor-universal coupling,
\begin{equation}
\label{eq:2loop_aff}
\mathcal{L} \supset  \frac{\alpha_{1,2}^2}{16 \pi^2 f_\phi} {\rm min}\left(1, \frac{y_L^2 f_\phi^2}{m_L^2} \right) \times \partial_\mu a \bar f \gamma^5 \gamma^\mu  f.
\end{equation}
The current bound on the axion-electron coupling from observations of white dwarf stars~\cite{Bertolami:2014wua} leads to an upper bound on $y_L$.
Two-loop corrections involving gluons (present if vector-like quarks are introduced) only induce vector couplings with quarks, which may be removed by baryon number rotations.

After integrating out $L$ and $\bar{L}$, the axion couples to $\ell$,
\begin{align}
\label{eq:a_ell_ell}
    \mathcal{L} \supset \frac{y_L^2 f_\phi^2}{2m_L^2}\frac{\partial_\mu a}{f_\phi}\ell^\dag \bar{\sigma}^\mu \ell.
\end{align}
In general, the coupling may violate the lepton flavor symmetry and induce rare processes such as $\mu \rightarrow e a$~\cite{Wilczek:1982rv,Calibbi:2020jvd}. Also, the axion-electron coupling is constrained by the observations of white dwarf stars~\cite{Bertolami:2014wua}.
A muon-electron-photon coupling is also generated by the quantum correction from $L$ and weak gauge bosons,
\begin{align}
\label{eq:mu_e_gamma}
    \mathcal{L} \supset \frac{ e g_2^2}{16 \pi^2} \frac{y_L^2 f_\phi^2 m_\mu}{m_L^4} F_{\mu\nu} \bar{\mu} \gamma^{\mu \nu} e + {\rm h.c.},
\end{align}
inducing $\mu \rightarrow e\gamma$.
These bounds can also be evaded if $L$ dominantly couples to the tau leptons. For this reason, we do not include these constraints in the figure but comment below on how they would impact the parameter space.

In summary, the green regions in Fig.~\ref{fig:model} are excluded because of the conflict between the thermalization constraint given in Eq.~(\ref{eq:yLmin_therm}) and the constraint from the Q-balls in Eq.~(\ref{eq:Q_ball}). These green regions are insensitive to the quantum correction constraint from Eq.~(\ref{eq:Delta_mS2}) as long as $\tilde m_{\widetilde{L}} < 4 \pi m_L$. The axino mass constraint in Eq.~(\ref{eq:AxinoMass_gaugino}) turns out to be irrelevant since  a tiny $y_L$ is sufficient for thermalization based on  Eq.~(\ref{eq:yLmin_therm}). We now comment on the constraints in Eqs.~(\ref{eq:a_ell_ell}) and (\ref{eq:mu_e_gamma}).  In the absence of any flavor texture, 1) the top panels would remain unaffected, and 2) in the bottom right panel, $f_\phi \gtrsim 10^7 \GeV$ becomes excluded by additional green regions.

We briefly describe the allowed range of $y_L$ in Fig.~\ref{fig:model}. The lower bound on $y_L$ originates from Eq.~(\ref{eq:yLmin_therm}). For the top panels, the bound is $y_L > 2 \times 10^{-7}$. For the bottom left (right) panel, the lower bound is $y_L > 3 \times 10^{-7}$ ($3 \times 10^{-6}$). The upper bound, by construction, is equal to the lower bound on the green boundary. For the top panels, the strongest upper bound comes from Eq.~(\ref{eq:Q_ball}) and scales as $y_L < 2 \times 10^{-7} (m_S / 200 \keV)$. For the bottom panels, the strongest upper bound comes from both Eqs.~(\ref{eq:Q_ball}) and (\ref{eq:AxinoMass_yL}) and thus depends on both $m_S$ and $f_\phi$ while staying under $4 \times 10^{-3}$ ($6 \times 10^{-5}$) throughout the allowed region for the left (right) panel.

\subsection{A baryogenesis connection}
\label{sec:baryogenesis}
The dynamics proposed in this paper may have intriguing connections with baryogenesis.
The charge in the rotation of the axion field can be transferred into the particle-antiparticle asymmetry in the thermal bath, allowing an explanation of the baryon asymmetry~\cite{Co:2019wyp,Co:2020xlh}.

In the concrete model studied in Sec.~\ref{sec:parameter_space}, the charge in the rotation is transferred into the chiral asymmetry of Standard Model particles via the coupling in Eq.~(\ref{eq:super_vector}), which is transferred into the baryon asymmetry via the electroweak sphaleron processes.
Using the result in Ref.~\cite{Co:2020xlh}, the baryon yield is given by 
\begin{equation}
\label{eq:YB_theta}
Y_B^\theta \simeq \left. \frac{45 c_B}{2 g_* \pi^2} \frac{\dot{\theta}}{T} \right|_{T_{\rm EW}}, \hspace{1cm} c_B \simeq \frac{25}{237} \times \min\left( 1, \frac{y_L^2 S^2}{2m_L^2} \right),
\end{equation}
where $T_{\rm EW}$ is the temperature when the electroweak sphaleron processes fall out of thermal equilibrium. The Standard Model predicts $T_{\rm EW} = 130 \GeV$~\cite{DOnofrio:2014rug}.

For the parameter choices relevant to Fig.~\ref{fig:model}, we find that the Yukawa coupling needed to reproduce the observed baryon asymmetry, $Y_B^{\rm obs} = 8.7 \times 10^{-11}$~\cite{Aghanim:2018eyx}, is given by
\begin{equation}
\label{eq:yL_YB}
    y_L \simeq 10^{-4} 
    \left( \frac{1}{r_\theta} \right)^{ \scalebox{1.01}{$\frac{1}{2}$} }
    \left( \frac{\rm mHz}{f_{\rm GW}} \right)^{ \scalebox{1.01}{$\frac{1}{2}$} }
    \left( \frac{10^{-12}}{\Omega_{\rm GW} h^2} \right)^{ \scalebox{1.01}{$\frac{1}{4}$} }
    \left( \frac{m_S}{\rm GeV} \right)^{ \scalebox{1.01}{$\frac{1}{2}$} }
    \left( \frac{m_L}{\rm TeV} \right)
    \left( \frac{130 \GeV}{T_{\rm EW}} \right) \left(\frac{200}{g_*(T_p)} \right)^{ \scalebox{1.01}{$\frac{1}{2}$} } .
\end{equation}
In the regions to the left of/above the cyan lines of the bottom panels in Fig.~\ref{fig:model}, the baryon asymmetry is underproduced because the required $y_L$ in Eq.~(\ref{eq:yL_YB}) exceeds some of the upper bounds on $y_L$ discussed in Sec.~\ref{sec:parameter_space}. Specifically, the vertical segments in both bottom panels are due to a conflict with the constraint from Q-balls, Eq.~(\ref{eq:Q_ball}). In the bottom right panel, the steeper sloped segment is set by the value of $y_L$ that saturates the axino mass constraint from Eq.~(\ref{eq:AxinoMass_gaugino}). Above the shallower sloped segment, $c_{B}$ is set by its saturated value in Eq.~(\ref{eq:yL_YB}), and for large $f_{\phi}$, the saxion soon settles to its minimum after tachyonic instability, and $\dot{\theta}$ is redshifted too much by the time of the electroweak phase transition to generate sufficient baryon number.   
In other words, even ignoring the bound from the axino mass, the maximum $f_{\phi}$ that allows sufficient production of the baryon asymmetry only changes by $\mathcal{O}(1)$.  
The regions of underproduction expand if $r_\theta < 1$. In the top panels of Fig.~\ref{fig:model}, the baryon asymmetry is always underproduced in the white regions.

We note that relaxing the constraint shown by the orange shading expands the parameter space compatible with the baryogenesis scenario. In fact, for $m_S>$ TeV, we may take the axino mass $\sim m_S$ above the LOSP mass, so that the axino decays into the LOSP via the coupling $y_L$ as discussed in Sec.~\ref{sec:potential}. In this case, only a milder constraint $m_S < f_\phi$ is applicable, actually opening up a parameter space with $m_S > $ TeV in the lower two panels of Fig.~\ref{fig:model}. This is not possible in the upper two panels.

Finally, in the absence of flavor structure in $y_L$, 
the cyan lines may be impacted.  This is because the flavor dependent constraints of the last section impose stronger upper bounds on $y_L$, which in turn impacts the baryon asymmetry via Eq.~(\ref{eq:yL_YB}).    Once these bounds are imposed, it is no longer possible to produce the full baryon asymmetry at all in the bottom right panel. In the bottom left panel, values of large $f_{\phi} \times m_S \; (\gtrsim 10^4~{\rm GeV}^2)$ that previously were able to reproduce the baryon asymmetry can no longer do so without running afoul of these  constraints. 

\section{Summary and Discussion}
\label{sec:discussion}

Axion fields frequently appear in extensions of the Standard Model that involve spontaneous breaking of global symmetries. Due to their small masses, axion fields in general do not rest at the minimum of the potential in the early universe. Rather, they are expected to be dynamical and, as a result, may play meaningful roles in the cosmological evolution of the universe. In this paper, we investigated consequences of an axion rotating in field space. Such a rotation is expected when the radial direction of the global symmetry breaking field takes on a large initial field value and the global symmetry is explicitly broken by higher dimensional operators.

When a rotating axion couples to a dark photon, it modifies the dispersion relation of the dark photon, and the dark photon can become tachyonic for a particular range of momenta. Dark photons are then efficiently produced by tachyonic instability. The produced dark photons generate gravitational waves whose frequencies may be of interest to current and future GW detectors such as NANOGrav, SKA, LISA, DECIGO, and BBO.

The produced dark photon may be dark matter.
The strength of the gravitational waves is correlated with the warmness of dark photon dark matter that can be probed by galaxy surveys and observations of 21-cm lines. The scattering among dark photons and axions can change the warmness, which should be investigated via a numerical lattice simulation.

Since this mechanism most naturally occurs in cases where the radial direction has a large initial field value, it is important to also consider the dynamics of this saxion.
The radial motion should be dissipated since otherwise it eventually dominates the energy density of the universe and leads to a moduli problem or dilution of gravitational waves.
We investigated the thermalization process in detail for a model where the global symmetry breaking field couples to new particles charged under the Standard Model gauge symmetry. In this case, thermalization is successful for a variety of GW signals, spanning a broad range of frequencies and strengths, including those that can explain the NANOGrav signal or be probed by future detectors such as SKA, LISA, DECIGO, and BBO.

The angular momentum of the rotation is transferred into particle-antiparticle asymmetry in the thermal bath through the sector responsible for the thermalization, which is subsequently transferred to a baryon asymmetry via electroweak sphaleron processes. The parameter space that explains the observed baryon asymmetry is also consistent with the thermalization requirement for a parameter region with high-frequency gravitational waves.

In this paper, we assumed that the axion is massless and becomes dark radiation. With a non-zero mass, the axion that induces tachyonic instability may itself be dark matter.  One possibility is that the axion is produced via the conventional misalignment mechanism~\cite{Preskill:1982cy,Abbott:1982af,Dine:1982ah}.  However, peculiar to the mechanism we consider, there are other possible axion dark matter production channels, where the axions are created with non-zero momenta and may be warm in contrast to the conventional misalignment mechanism. We leave detailed discussion of this possibility for future work, but here we outline three interesting possibilities:
\begin{enumerate}[leftmargin=*]
\item The kinetic misalignment mechanism~\cite{Co:2019jts,Co:2020dya}, where the axion energy originates from the kinetic energy of the axion rotation. In Refs.~\cite{Co:2019jts,Co:2020dya}, the axion rotation was assumed to remain coherent and oscillates when $\dot{\theta}\sim m_\phi$. However, as pointed out in Refs.~\cite{Jaeckel:2016qjp, Berges:2019dgr} and Ref.~\cite{Fonseca:2019ypl} in the context of the axion monodromy dark matter and relaxion respectively, parametric resonance arising from the axion self-interactions is important.  The coherent axion motion turns into axion fluctuations at a rate $\sim m_\phi^4/\dot{\theta}^3$ with a momentum $\simeq \dot{\theta}/2$~\cite{Fonseca:2019ypl}. This occurs before the oscillations begin. Since the energy of the axion per quantum $\simeq \dot{\theta}/2$, the yield of the axions $Y_{\phi} \simeq \rho_\theta/(\dot{\theta}s/2) = Y_\theta = r_\theta Y_{\theta,i}$.%
\footnote{
This estimation is the same as that in Ref.~\cite{Co:2019jts} up to a factor of two. This is because the axion energy per quantum is also around $\dot{\theta}$  when axions are produced as coherent oscillations neglecting the growth of axion fluctuations.
}
The axions may be too warm for a small decay constant and/or a small axion mass, where the growth of fluctuations occurs at a late time, which gives the axion momentum an inadequate opportunity to redshift. Estimation of the axion abundance from the kinetic misalignment mechanism requires a more detailed estimation of the fraction of the charge that remains in rotation after tachyonic instability, denoted by $r_\theta$ in Eq.~(\ref{eq:rtheta}).  

\item Parametric resonance from the radial oscillation mode~\cite{Co:2017mop,Co:2020dya,Co:2020jtv}, where the axion energy originates from the potential and kinetic energy of the radial mode. Since the energy of the axions per quantum is determined by the frequency of the zero mode motion, $m_S\sim \dot{\theta}$, for a motion with ${\mathcal O}(1)$ ellipticity, the yield of axions $Y_{\phi} \sim \rho_{\theta,i}/(\dot{\theta}s)\sim Y_{\theta,i}$. If $r_\theta \ll 1$, this contribution dominates over that of the kinetic misalignment mechanism.

\item The production of axion fluctuations from the scattering of the dark photons produced by tachyonic instability.  
Since the momentum of the produced axion $\sim k_{\rm TI} \ll \dot{\theta}$, if $\mathcal{O}(1)$ fraction of the energy of the axion rotation is eventually transferred into the axion fluctuations in this way, the number density of axions produced in this channel, $Y_\phi \sim \rho_{\theta,i}/(\omega(k_{\rm TI})s)$, is much larger than those in 1) and 2). Here $\omega(k_{\rm TI})$ is the energy of an axion quantum with a momentum $k_{\rm TI}$ that depends on the shape of the saxion potential. Moreover, the scattering among axions and dark photons, as argued in Sec.~\ref{sec:DPDM_scatt}, can change the spectrum of axions.
\end{enumerate}

This work motivates future numerical studies.
In this mechanism, the energy of the axion rotation is transferred into that of dark photons. We expect that the backreaction to the axion rotation stops the transfer, and some amount of energy remains in the axion rotation. Here we simply treated the residual amount of energy as a free parameter, but this should eventually be fixed by numerical lattice computations. These computations will also reveal the abundance of axions produced in the channels 1) and 3) discussed in the previous paragraph. This will allow the investigation of the interesting possibility of identifying the axion we discuss in this paper with the QCD axion.
Also, we estimated the amount and peak frequency of the produced dark photons and gravitational waves analytically. However, a numerical lattice computation could precisely determine the spectral shape.  In principle, this could then be contrasted with those of other production mechanisms. Such a simulation would also allow a more precise prediction on the correlation between GW signals and the warmness of dark photon dark matter or the amount of dark photon dark radiation.

\section*{Acknowledgments} 
A.P. would like to thank T.~Opferkuch for a useful discussion. R.C. and K.H. would like to thank Ryosuke Sato, Ken Van Tilburg, and Yue Zhao for bringing our attention to the growth of axion fluctuations that was discussed in Refs.~\cite{Jaeckel:2016qjp, Berges:2019dgr, Fonseca:2019ypl}.  The work was supported in part by the DoE Early Career Grant DE-SC0019225 (R.C.) and DoE grant DE-SC0011842 at the University of Minnesota (R.C.), the DoE grant DE-SC0007859 (A.P.), and Friends of the Institute for Advanced Study (K.H.).  A.P. would also like to thank the Simons Foundation for support during his sabbatical.

\appendix

\renewcommand{\thesubsection}{\thesection.\arabic{subsection}}

\section{Cosmological Details of Concrete Thermalization Model}
\label{app:concrete_cosmo}

In this Appendix, we describe the cosmological fate of new particles beyond the Standard Model.
We find that all new particles will either be dark matter, comprise harmless dark radiation, or decay before BBN or after it invisibly.

First, we consider gravitinos. Requiring their abundance not to be so large as to create a gravitino-dominated era bounds the reheat temperature.  This temperature sets a starting point for understanding the rest of the cosmology.  In particular, we will require this reheat temperature be larger than other scales relevant for the cosmology, including the temperature at the time of the tachyonic instability, $T_{p}$.

 We expect that the saxion mass and gravitino mass are comparable, as we imagine that supersymmetry breaking is mediated to the $P$ sector via Planck-suppressed operators. As a reminder, for the case of a NANOGrav or SKA signal in the top panels of Fig.~\ref{fig:model}, we expect a gravitino mass in the range of $\mathcal{O}(10 \keV)$ to $\mathcal{O}({\rm MeV})$, whereas for LISA and DECIGO/BBO in the bottom panels, we find gravitino masses in the range of $\mathcal{O}({\rm MeV})$ to $\mathcal{O}({\rm TeV})$.

\begin{enumerate}[leftmargin=*]

\item The gravitinos are produced during inflationary reheating and later decay to the axinos.
The gravitinos are thermally produced from the scattering of Minimal Supersymmetric Standard Model (MSSM) particles such as gauge bosons and gauginos~\cite{Nanopoulos:1983up}. When $m_{3/2}$ is much smaller than the gaugino masses, the yield of thermal production computed in Refs.~\cite{Moroi:1993mb, Rychkov:2007uq} can be approximated as
\begin{align}
\label{eq:Ygravitino}
    Y_{3/2} \sim 10^{-6} \left( \frac{T_R}{10^{10} \GeV} \right) \left( \frac{\rm GeV}{m_{3/2}} \right)^2 \left( \frac{m_{\widetilde g}}{\rm TeV} \right)^2 \times \left\{\begin{array}{ll}
1  & \quad \text { for } \quad T_R > m_{\widetilde g} \\
\left( \frac{T_R}{m_{\widetilde g}} \right)^6  & \quad \text { for } \quad T_R < m_{\widetilde g}
\end{array}\right. ,
\end{align}
where $T_R$ is the reheat temperature after inflation.  Here, $m_{\widetilde g}$ is the gluino mass. The $A$-term associated with that top Yukawa coupling is assumed to be of the same order as the gaugino masses, for which case the contribution from the $A$-term is negligible.  The expression for $T_R < m_{\widetilde g}$ is correct for a sufficiently heavy inflaton, where the non-thermal production is subdominant~\cite{Harigaya:2014waa, Harigaya:2019tzu}. We also caution that Eq.~(\ref{eq:Ygravitino}) breaks down when $T_R \ll m_{\widetilde g}$; gravitino production is typically dominated by processes at lower temperatures during inflationary reheating, and therefore the contribution from lighter binos can dominate over that of the gluinos despite a smaller gauge coupling constant. 

If the gravitino decays to the axino and the axion before dominating the energy density, the contribution to dark radiation is negligible.
For $m_{3/2} \lesssim 10 \GeV$, the decay occurs after matter-radiation equality and hence we require that the gravitino abundance be less than that of dark matter. For $m_{3/2} \gtrsim 10 \GeV$, the decay occurs before matter-radiation equality and the gravitino energy density at this time should be below that of radiation.
We obtain an upper bound on the reheat temperature from Eq.~(\ref{eq:Ygravitino}),
\begin{align}
\label{eq:TR_gravitino}
T_R & \lesssim 
\left\{\begin{array}{ll}
4 \times 10^6 \GeV \left( \frac{m_{3/2}}{\rm GeV} \right) \left( \frac{\rm TeV}{m_{\widetilde g}} \right)^2 \\
6 \times 10^2 \GeV \left( \frac{m_{3/2}}{10 \keV} \right)^{ \frac{1}{7} } \left( \frac{m_{\widetilde g}}{\rm TeV} \right)^{ \frac{4}{7} } \\
4 \times 10^7 \GeV \left( \frac{m_{3/2}}{10 \GeV} \right)^{\frac{5}{2}} \left( \frac{\rm TeV}{m_{\widetilde g}} \right)^2 \\
5 \times 10^3 \GeV \left( \frac{m_{3/2}}{10 \GeV} \right)^{ \frac{5}{14} } \left( \frac{m_{\widetilde g}}{\rm TeV} \right)^{ \frac{4}{7} }
\end{array} \right.
& \begin{array}{ll}
\text { for } \ \  T_R > m_{\widetilde g} \\
\text { for } \ \  T_R < m_{\widetilde g} \\
\text { for } \ \  T_R > m_{\widetilde g} \\
\text { for } \ \  T_R < m_{\widetilde g}
\end{array} 
\begin{array}{ll}
\left\} \begin{array}{ll}
\\  \\
\end{array} \right. m_{3/2} < 10 \GeV
\\
\left\} \begin{array}{ll}
\\  \\
\end{array} \right. m_{3/2} > 10 \GeV
\end{array} .
\end{align}
Noting that $m_{3/2} \sim m_S$, this equation indicates that $T_R$ can be chosen consistently higher than $T_p$ in Eq.~(\ref{eq:Tprod}) for all of the parameter space shown in Fig.~\ref{fig:model}. For reference, the values of $T_{p}$ for the benchmarks shown in the figure are enumerated below Eq.~(\ref{eq:rhorot}).  Thus, the assumption of a radiation-dominated universe during thermalization is satisfied.

\item The axions contribute to dark radiation. In this paper, we assume that the axion is massless (although the axion may be dark matter via different production channels described in Sec.~\ref{sec:discussion}). In the case of massless axions, the remaining energy density in the coherent rotation $\rho_\theta$ scales as matter, $\rho_\theta \propto R^{-3}$, when $S > f_\phi$ and as kination, $\rho_\theta \propto R^{-6}$, after $S \simeq f_\phi$. This implies that $\rho_\theta$ eventually redshifts away and becomes harmless as long as it does not change the energy budget around BBN and the decoupling of the CMB. In addition to this zero-mode contribution,  thermalized axions and the axions from saxion decays both redshift as radiation and contribute to dark radiation; relevant discussions are around Eqs.~(\ref{eq:DeltaNeff}) and (\ref{eq:s_aa}).

\item The axino is assumed to be the lightest supersymmetric partner (LSP) and constitutes a small contribution to dark radiation.  If the axino is lighter than $\mathcal{O}({\rm eV})$~\cite{Osato:2016ixc}, even the thermal contribution of the axino is harmless.  We impose this constraint, noting that axinos decouple from the bath when the fields responsible for thermalization leave the bath.
Since we assume the axino is the LSP, other $R$-parity odd particles will eventually decay to them, as discussed above for the gravitino. For other $R$-parity odd particles, we will verify below that their decay occurs when their energy densities are subdominant compared to radiation, and thus do not produce excessive dark radiation. 

\item The dark photino mass can be as light as $m_{\widetilde A'} \simeq e_D^2 m_{3/2} / 16\pi^2$ from the quantum correction in anomaly mediation~\cite{Giudice:1998xp,Randall:1998uk} (see also~\cite{Bagger:1999rd,Boyda:2001nh,Harigaya:2014sfa}).%
\footnote{The threshold correction from the dark fermion-sfermion loops is negligible and does not cancel the UV anomaly mediation. This is because in the model with dimensional transmutation, the $B \mu$ term of $\psi\bar{\psi}$ is only given by anomaly mediation and is much smaller than $y_\psi f_\phi m_{3/2}$.}
From Eq.~(\ref{eq:gaugecoupling}) and taking $m_{3/2} \simeq m_S$, we note that the photino may be quite light,
\begin{equation}
    \label{eq:photinomass}
    m_{\widetilde A'} \simeq 1  \keV  \left( \frac{f_{\rm GW}}{0.1~\rm{Hz}} \right)^2 \left( \frac{20}{r_p} \right) \left(\frac{g_*(T_p)}{200} \right)^{ \scalebox{1.01}{$\frac{1}{6}$} }.
\end{equation}
Even for the highest frequency we consider (the DECIGO/BBO benchmark of $f_{\rm GW} =$ 0.1 Hz), this modest mass implies that as long as the dark photino number density is smaller than one-tenth of the thermal one, the dark  photino energy density is negligible. Dark photinos are produced by annihilation of axions, saxions, axinos, or vector-like fermions, but since the production rate is suppressed by the decay constant, a loop factor, and the dark gauge coupling, we find that the dark photino number density is much smaller than the thermal number density in the entire viable parameter space.

\item The dark photons are either dark radiation or dark matter depending on the mass.
Dark photons are copiously produced by tachyonic instability as discussed in Sec.~\ref{sec:TI}. The dark photons may be cold enough to serve as dark matter, depending on the dark photon mass and the GW signal as we elaborate in Sec.~\ref{sec:DPDM} and Fig.~\ref{fig:GW_DPDM}. 

If the mass is smaller than that required for dark matter, dark photons behave as dark radiation with contribution to $\Delta N_{\rm eff}$ shown in Eq.~(\ref{eq:DeltaNeff}), which can be significant for large GW amplitudes, like a putative NANOGrav signal. In this case, subsequent thermalization of the dark photons by the saxion and particles charged under the global symmetry 
does not introduce additional $\Delta N_{\rm eff}$ if these particles are already decoupled from the Standard Model bath.

On the other hand, thermalization would be disastrous for dark photon dark matter--it will necessarily heat dark photons while reducing its number density.
This can occur through the scattering with $\psi$, saxions/axions, or $L$ in the thermal bath. We find that we may always take $y_\psi$ large enough so that the mass of $\psi\sim y_\psi S$ is much larger than the temperature and $\psi$ is never in the thermal bath. The scattering rate with thermal saxions/axions is $\sim (\alpha_D/4\pi)^2 k_{A'}T^4/S^4$, where $k_{A'}$ is the momentum of dark photons. The rate turns out be always smaller than the Hubble expansion rate. The scattering rate with $L$ is$\sim (\alpha_D/4\pi)^2 y_L^2 k _{A'}T^2/S^2$. After requiring $y_L S < T$, the rate is smaller than the rate with saxions/axions.

\item The fermion $\psi$ charged under $U(1)_P$ and the dark gauge symmetry (and their superpartners) is kinematically inaccessible and thus never populated. In a small fraction of the parameter space, it is also possible that $\psi$ are populated after tachyonic instability and the freeze-out abundance of it explains the observed dark matter abundance.

We recall that the coupling to the dark photons was induced via the presence of the fermions $\psi$ charged under the dark gauge symmetry, with an anomaly under the global symmetry, see Eq.~(\ref{eq:mediator}).  At $T_{p}$, when the tachyonic instability takes place, the $U(1)_P$-charged particles must be absent from the thermal bath to avoid excessive electric conductivity that prevents efficient tachyonic production. As discussed in Sec.~\ref{sec:various}, we find that a proper choice of $y_\psi < 1$ allows the fermion mass to be higher than the temperature at $T_p$, thereby suppressing their production during this crucial era. 

However, since their mass depends on the $S$ field value and redshifts, these $U(1)_P$-charged particles could, at least in principle, come into thermal equilibrium at a later time.  We have found that there are choices of $y_{\psi}$, consistent with the bound on the axino mass Eq.~(\ref{eq:axinomassnumerical}) such that this does not occur so that these particles are never produced.

Nevertheless, in the case where $y_{\psi}$ is small enough that production is not always forbidden, it is an interesting question whether they could be the dark matter.
The $U(1)_P$-charged fermions $\psi$ dominantly annihilate into the axions by $t$- or $u$-channel exchange of $\psi$ with an approximate cross section 
\begin{equation}
\sigma v 
(\psi \bar{\psi} \rightarrow aa) \simeq \frac{m_\psi^2 }{8\pi f_\phi^4}.
\end{equation}
We first discuss the case of the NANOGrav or SKA signal where the dark photon is not a viable dark matter candidate due to warmness--unless a cascade towards IR occurs due to scattering--and an alternate dark matter candidate would be especially welcome. To have sufficiently cold dark matter, $m_\psi \gtrsim 10$ keV is required~\cite{Yeche:2017upn}, which requires $f_\phi \gtrsim 0.1$ GeV to achieve the correct freeze-out abundance.
Here we borrow the warmness constraint on sterile neutrino dark matter produced from oscillations of active neutrinos, since the momentum distribution for that case is expected to be similar to the freeze-out case.
As can be seen from Fig.~\ref{fig:model}, in the region with the thermalization constraint, $f_\phi \sim 0.1-1$ GeV.  Then, achieving the right freeze-out density would require $m_\psi\sim 10$ keV.
Since $S_p \sim f_\phi$ in that parameter region, $m_\psi = y_\psi S$ is necessarily smaller than $T_p \simeq 5.5 \MeV (27 \MeV)$ for NANOGrav (SKA) and is already produced from the thermal bath at the time of GW production, leading to excessive electric conductivity and hence spoiling the tachyonic instability.  Therefore, $\psi$ cannot be the dark matter in this case.
 
While in most of the parameter space in the bottom panels of Fig.~\ref{fig:model}, $\psi$ is too abundant if its production is kinematically allowed, around the region $10 \MeV < f_\phi \lesssim 10 \GeV$ of the bottom left panel that is excluded by BBN if $r_{\theta} > \mathcal{O}(0.1)$, the freeze-out abundance of $\psi$ can explain the observed amount of dark matter while $\psi$ is cold enough.

When the motion of $P$ is not completely circular, the mass of $\psi$ oscillates and $\psi$ may be produced by parametric resonance. However, as long as $y_\psi S$ is much larger than $m_S$, which is the case in the viable parameter space, the oscillation of the mass is adiabatic and the production by parametric resonance is exponentially suppressed. 

\item The vector-like fermions decay into a Standard Model doublet lepton and a saxion/axion via the Yukawa coupling $y_L$ with a rate $\sim y_L^2 m_L/ (8\pi)$ and the decay occurs much before BBN. The vector-like fermions mix with Standard Model doublet leptons with an angle $\sim y_L f_\phi/m_L$. This mixing causes them to decay into a Higgs boson and a singlet lepton, which we assume is a tau, with a rate $\sim y_L^2 y_\tau^2 f_\phi^2/m_L$.  These decays occur before BBN for the parameter region in Fig.~\ref{fig:model}.

\item The vector-like sfermions decay into the vector-like fermions and gauginos. With the aforementioned Yukawa coupling or mixing, if the masses of vector-like sfermions are larger than those of MSSM sfermions, the vector-like sfermions decay into MSSM sfermions. The case of the vector-like sfermion LOSP is discussed later.

\item 
The LOSPs freeze out and decay to axinos or gravitinos and the Standard Model particles before BBN.

For $m_{3/2} < {\rm GeV} (m_{\rm LOSP}/{\rm TeV})^{5/2}$, the decay of the LOSP into a gravitino occurs before BBN and the resulting gravitino abundance may be below the dark matter abundance~\cite{Feng:2004zu,Feng:2012rn} so that the subsequent decay of the gravitino into an axino does not overproduce dark radiation however slow the decay is. 

For a larger gravitino mass, the direct decay of the LOSP into an axino and Standard Model particles should  occur before BBN.
In fact, the LOSPs can decay via the coupling $y_L$ of new vector-like fermions with $P$. For example, if the LOSP is the bino and $m_L < m_{\widetilde{B}}$,
\begin{align}
\label{eq:bino_LOSP}
    \Gamma(\widetilde{B}\rightarrow \ell \bar{L} \widetilde{a}) \simeq \frac{g'^{2} y_L^2}{128\pi^3} m_{\widetilde{B}} \simeq 10^{-16} \GeV \left( \frac{m_{\widetilde{B}}}{\rm TeV} \right) \left( \frac{y_L}{10^{-7}}\right)^2.
\end{align}
In the benchmark points in Fig.~\ref{fig:model}, $y_L \gtrsim 10^{-7}$ is required by Eq.~(\ref{eq:yLmin_therm}), so the decay occurs well before BBN. The vector-like leptons then decay promptly as discussed above. If $m_L >m_{\widetilde{B}}$, decay occurs through the K\"ahler potential $y_L^2 PP^\dag \ell \ell^\dag / m_L^2$. The decay rate is suppressed by the mass of the fermion component of $\ell$, which we assume to be $\tau$,
\begin{align}
    \Gamma(\widetilde{B} \rightarrow \ell \ell^\dag \widetilde{a} S/a) \simeq & \frac{{g'}^2 y_L^4}{2048 \pi^4} \frac{m_{\widetilde{B}}^3 m_\tau^2}{m_L^4} \simeq 10^{-21}~{\rm GeV} \left( \frac{y_L}{10^{-3}}\right)^4 \left(\frac{m_{\widetilde{B}}}{\rm TeV}\right)^3 \left( \frac{\rm TeV}{m_L}\right)^4,  \\
    \Gamma(\widetilde{B}\rightarrow \ell \ell^\dag \widetilde{a}) \simeq & \frac{{g'}^2 y_L^4}{128 \pi^3} \frac{f_\phi^2 m_{\widetilde{B}}m_\tau^2}{m_L^4} \simeq 10^{-21}~{\rm GeV} \left( \frac{y_L}{10^{-5}}\right)^4 \left( \frac{m_{\widetilde{B}}}{\rm TeV} \right) \left(\frac{\rm TeV}{m_L}\right)^4 \left(\frac{f_\phi}{10^6 \GeV}\right)^2, \nonumber
\end{align}
where we assume $m_{\widetilde{B}} \sim m_{\widetilde{\ell}}$.
To ensure decay before BBN requires large enough $y_L$ and/or $f_\phi$ that are inconsistent with the low-frequency parameter regions, but the decay into a gravitino anyway occurs before BBN because of the small $m_S\sim m_{3/2}$.

If the LOSP is the superpartner of the new fermion,  $\widetilde{L}$, it decays into $\ell +\widetilde{a}$ with a rate $\sim y_L^2 m_{\widetilde{L}}/(8\pi)$, which is much larger than in Eq.~(\ref{eq:bino_LOSP}).

If the LOSP is the right-handed stau $\widetilde{\bar{\tau}}$ and $m_{\widetilde{\bar{\tau}}}> m_L$,
the stau decays into $L+\widetilde{a}$ with a rate
\begin{align}
    \Gamma ( \widetilde{\bar{\tau}}\rightarrow \bar{L} \widetilde{a}) \simeq \frac{y_L^2}{8\pi}  {\rm tan}^2\beta \frac{m_\tau^2 \mu^2}{m_{\widetilde{\bar{\tau}}}^3} \simeq 10^{-16}~{\rm GeV} \left( \frac{y_L}{10^{-7}}\right)^2 \left( \frac{{\rm tan}\beta}{10}\right)^2 \left( \frac{\mu}{\rm TeV}\right)^2 \left( \frac{\rm TeV} {m_{\widetilde{\bar{\tau}}}}\right)^3  ,
\end{align}
which occurs before BBN.
If $m_{\widetilde{\tau}} < m_L$,
the decay occurs via the aforementioned K\"ahler potential. Taking the $F$ term of $\ell$ and the fermion components of $\ell^\dag$ and $P^\dag$, the decay rate is given by
\begin{align}
    \Gamma(\widetilde{\bar{\tau}}\rightarrow  \ell^\dag \widetilde{a} S/a) \simeq & \frac{y_L^4}{128\pi^2}  \frac{m_\tau^2 m_{\widetilde{\bar{\tau}}}^3}{m_L^4} \simeq 10^{-22}\GeV \left( \frac{y_L}{10^{-4}}\right)^4 \left(\frac{m_{\widetilde{\bar{\tau}}}}{\rm TeV}\right)^3 \left( \frac{\rm TeV}{m_L}\right)^4,  \\
    \Gamma(\widetilde{\bar{\tau}}\rightarrow  \ell^\dag \widetilde{a}) \simeq & \frac{y_L^4}{8\pi}  \frac{m_\tau^2 f_\phi^2 m_{\widetilde{\bar{\tau}}}}{m_L^4} \simeq 10^{-22}\GeV \left( \frac{y_L}{10^{-6}}\right)^4 \left( \frac{m_{\widetilde{\bar{\tau}}}}{\rm TeV} \right) \left( \frac{\rm TeV}{m_L}\right)^4 \left(\frac{f_\phi}{10^6 \GeV}\right)^2. \nonumber
\end{align}
The decay before BBN requires large enough $y_L$ and/or $f_\phi$.

\end{enumerate}

\bibliography{GW_rotation}

\end{document}